\documentclass[12pt,preprint]{aastex}
\usepackage{emulateapj5}

\newcommand{\zform}{z_{\rm form}}
\newcommand{\omstars}{\Omega_{\rm stars}}

\newcommand{\bj}{$b_{\rm J}$}

\newcommand{\Msun}{\hbox{M$_{\odot}$}}
\newcommand{\Lsun}{\hbox{L$_{\odot}$}}
\newcommand{\Zsun}{\hbox{Z$_{\odot}$}}
\newcommand{\sfrunits}{$h$ $\Msun$\,yr$^{-1}$ Mpc$^{-3}$}

\begin{document}

\title{The Sloan Digital Sky Survey: 
  The Cosmic Spectrum and Star-Formation History.}
\shorttitle{SDSS: the cosmic spectrum and SFH}

\author{{} Karl Glazebrook$^1$, Ivan K.\ Baldry$^1$, Michael R.\ 
  Blanton$^2$, Jon Brinkmann$^3$, Andrew Connolly$^4$, Istv\'an Csabai$^5$, Masataka
  Fukugita$^6$, \v{Z}eljko Ivezi\'{c}$^7$, Jon Loveday$^8$, Avery Meiksin$^9$, Robert Nichol$^{10}$, Eric Peng$^1$,
  Donald P. Schneider$^{11}$, Mark SubbaRao$^{12}$, Christy Tremonti$^1$, Donald G. York$^{12}$}
\shortauthors{K. Glazebrook et al.}

\altaffiltext{1}{Department of Physics \& Astronomy, Johns Hopkins
  University, Baltimore, MD~21218-2686, USA}
\altaffiltext{2}{Center for Cosmology and Particle Physics, Department
  of Physics, New York University, 4 Washington Place, New York, NY
  10003}
\altaffiltext{3}{Apache Point Observatory, P.O. Box 59, Sunspot, NM
  88349, USA}
\altaffiltext{4}{Department of Physics and Astronomy, University of Pittsburgh, 100 Allen Hall, 
  3941 O'Hara Street, Pittsburgh, PA 15260}
\altaffiltext{5}{Department of Physics of Complex Systems, 
		E\"otv\"os University, P\'azm\'any P\'eter s\'et\'any 1, 
		H-1518 Budapest, Hungary}
\altaffiltext{6}{Institute for Cosmic Ray Research, University of
  Tokyo, 5-1-5 Kashiwa, Kashiwa City, Chiba 277-8582, Japan}
\altaffiltext{7}{Princeton University Observatory, Peyton Hall, Princeton,NJ 08544-1001}
\altaffiltext{8}{Astronomy Centre, University of Sussex, Falmer,
  Brighton BN1 9QJ, UK}
\altaffiltext{9}{Institute for Astronomy, Royal Observatory,
  University of Edinburgh, Blackford Hill, Edinburgh EH9 3HJ, UK}
\altaffiltext{10}{Department of Physics, Carnegie Mellon University,
  5000 Forbes Avenue, Pittsburgh, PA 15232, USA}
\altaffiltext{11}{Department of Astronomy and Astrophysics, 525 Davey
  Laboratory, Pennsylvania State University, University Park, PA
  16802, USA}
\altaffiltext{12}{Department of Astronomy and Astrophysics, University
  of Chicago, 5640 South Ellis Avenue, Chicago, IL 60637, USA}

\begin{abstract}
  We present a determination of the `Cosmic Optical Spectrum' of the
  Universe, i.e.\ the ensemble emission from galaxies, as determined
  from the red-selected Sloan Digital Sky Survey main galaxy sample
  and compare with previous results of the blue-selected 2dF Galaxy
  Redshift Survey. Broadly we find good agreement in both the spectrum
  and the derived star-formation histories. If we use a power-law
  star-formation history model where 
  star-formation rate $\propto (1+z)^\beta$ out to $z=1$, then we find that
  $\beta$ of 2 to 3 is still the most likely model and there is no
  evidence for current surveys missing large amounts of star formation
  at high redshift. In particular `Fossil Cosmology' of the local
  universe gives measures of star-formation history which are
  consistent with direct observations at high redshift.  Using the
  photometry of SDSS we are able to derive the cosmic spectrum in
  absolute units (i.e.\ W \AA$^{-1}$ Mpc$^{-3}$) at 2--5\AA\ 
  resolution and find good agreement with published broad-band
  luminosity densities.  For a Salpeter IMF the best fit stellar mass/light ratio is
  3.7--7.5 $\Msun/\Lsun$ in the $r$-band (corresponding to $\omstars h
  = 0.0025$--0.0055) and from both the stellar emission history and
  the H$\alpha$ luminosity density independently we find a
  cosmological star-formation rate of 0.03--0.04 \sfrunits\ today.
\end{abstract}



\section{Introduction}
\label{sec:intro}


The comoving star-formation rate (SFR) of the Universe is in decline.
Since $z=1$ it has dropped by a factor of 3--15 \citep{CSB99,LLHC96}.
At higher redshifts it may have been constant or declining from $z=1$
to $z=5$ but the evidence is that $z=1$ is a critical point for the
universal average SFR \citep{madau96,steidel99}.

This is a very dramatic conclusion and worth tackling with a variety
of complementary observational techniques. One must of course consider
how SFR is measured at different redshifts: for $z>1$ the principal
observational probe has been rest-frame ultraviolet emission since it
is easily accessible in optical bandpasses. Because the UV derives
from young stellar populations ($t << 1$Gyr for $\lambda_{\rm rest}
\la 2000$\AA) the UV flux {\em produced} per galaxy will be
proportional to the SFR. There are details to do with the
contamination from older populations but for $z>1$ and $\lambda_{\rm
  rest} \la 2000$\AA\ these are minor \citep{MPD98}. A very important
issue though is dust extinction which can be many magnitudes in the UV
and in fact represents the principal uncertainty in whether the SFR
drops off again at high redshift \citep[see for
example][]{pettini98,steidel99}. A comprehensive discussion of UV dust
extinction issues is given by \cite{bell02}. Another important issue
is the treatment of cosmological surface-brightness dimming which may
result in surveys missing a dramatic increase in SFR at high redshift
\citep{Lanzetta02}.

At lower redshifts, where the rest-frame optical spectrum is
accessible, alternative methods of calculating SFRs from line emission
can be used. The simplest is H$\alpha$ which traces the number of
Lyman continuum photons \citep{glaze99,HCS00}. This technique has the
advantage of using radiation emitted at red wavelengths and is thus
considerably less sensitive to dust extinction compared to the UV.
Other lines such as H$\beta$ \citep{TM98} and [OII] \citep{HCBP98} are
used but their relationship to SFR is more complicated. For $z<1.5$
some of these lines are accessible in the optical and near-infrared
windows and can be used to measure SFRs. A review is given by
\cite{Hogg02} --- the vast majority of the surveys that have been
carried out indicate considerable evolution in the SFR for $z<1$
although there is disagreement on the amount of evolution required.

A complementary method has been to probe the far-infrared emission of
galaxies ($10\micron<\lambda<300\micron$) where dust-processed UV is
re-emitted thermally. At high redshifts this must be followed into the
sub-mm radio wavelengths. The emission has been used to constrain SFR
at low-redshift \citep{rowan97} and at high redshift \citep{hughes98}
but this approach suffers from both uncertainty in the dust modeling
and a lack of spectroscopic redshifts.  The latter issue is addressed
by fitting the spectral energy distributions (SED) with a series of
templates in order to photometrically estimate the redshifts, however
there are huge degeneracies between the photometric redshift estimate
and the assumed temperature of the dust SED \citep{blain02}.

All these approaches involve estimating a luminosity, either continuum
or line, per galaxy and then multiplying by the space density in order
to give a luminosity per comoving volume. At this point the scale
factor, SFR/luminosity, which is where the main uncertainties arise,
allows transformation to SFR per unit volume. The light budget per
volume is a useful quantity because it allows the stellar emission
history of the Universe to be decoupled, in a sense, from its
dynamical history (i.e.\ changes in the number of counted objects by
processes such as galaxy formation and galaxy-galaxy merging).  This
use of luminosity density is a {\em direct} method, in which an
observed luminosity density at a given redshift is converted to a SFR
density at the same redshift.

An alternative approach is that of fossil cosmology where the past
history of the Universe is determined from its current contents. This
can be done by examining the resolved stellar populations in the Local
Group \citep{HIC01} or in ensembles of galaxies, for example
in early type galaxies (e.g. \cite{Bernard02,eisenstein02}). Our approach is
to look at the ensemble of {\em all} galaxies; the `Cosmic Optical
Spectrum' of the local Universe. This represents the luminosity-scaled spectra
summed over all galaxies.  The cosmic spectrum can be thought of as the total
emission from all the objects in a representative volume of the
Universe.\footnote{In reality we can only measure galaxies down to
  some limiting magnitude, so any calculated cosmic spectrum is just
  an estimate of the true Cosmic Spectrum.} Objects contribute to the
cosmic spectrum according to their luminosity.  As in the case for an
individual galaxy, this spectrum contains a luminosity-weighted mix of
features from both old and young stars\footnote{There is of course AGN
  activity but this is a negligible contribution to the total optical
  emission as we will see later} and we can fit models of
star-formation history to it. In particular because the cosmic
spectrum represents an average, it will represent the end point of the
average SFH. Thus we can fit much simpler models to the cosmic spectrum
than are required for the spectra of individual galaxies, since we
expect the SFH history of the Universe, as a whole. to vary smoothly
with time. 

This ensemble approach was applied by \citeauthor{BAL02} (2002,
hereafter BG02) to the cosmic spectrum (meaning the optical spectrum
per unit volume) of 166\,000 galaxies in the 2dF Galaxy Redshift
Survey \citep[2dFGRS,][]{GRS01} and derived constraints on allowable
star-formation histories (SFH) which agreed well with results derived
from direct high redshift measurements via luminosity densities.

The Sloan Digital Sky Survey \citep[SDSS,][]{SDSS} provides many
advantages over the 2dFGRS for this type of analysis: it is of higher
spectral resolution (though we are not yet able to exploit this in
this paper) and will be about four times larger upon completion. The
spectral wavelength coverage is larger ($3600\rm\AA<\lambda<8000\AA$
for 2dFGRS and $3800\rm\AA<\lambda<9200\AA$ for SDSS). The photometric
calibration is much better as each SDSS multi-fiber observation
contains numerous standard stars and can be individually calibrated,
whereas for the 2dFGRS a mean calibration was applied to all the
survey spectra. The excellent spectrophotometry
is borne out by the good agreement between
synthetic colors computed from the spectra and actual colors which agree,
in average spectra to $<\pm 5$\% (Tremonti
2002, private communication). Also the accurate five-color photometry
allows comparisons of photometric constraints with color constraints.
In particular SDSS is selected in the $r$ band 
($\lambda = 6200\pm600$\AA), whereas 2dFGRS is selected in \bj\ 
($\lambda = 4600\pm 700$\AA). 
Thus one would expect SDSS to be more biased toward old
massive galaxies and 2dFGRS to be biased toward young, star-forming
galaxies. The comparison between the two allows us to investigate the
uncertainties in the determination of the SFH.

In this paper we compute the cosmic spectrum for the SDSS local
volume, we make a direct comparison with that derived from 2dFGRS by
BG02, and we derive new constraints on star-formation history models
from the SDSS spectrum.  The plan of this paper is as follows. In
Section~\ref{sec:data} we describe the SDSS data and our methods for
combining the spectra to form cosmic spectra.  In
Section~\ref{sec:results} we describe our modeling and fitting
procedure and the outcome of the comparison of best fitting SFHs
between the SDSS and 2dFGRS surveys.  We also test the consistency of
models of SFH which form a lot of stars at $z>1$. In
Section~\ref{sec:phys} we generate an absolute cosmic spectrum which
we show in physical units. We use this to estimate emission-line
luminosity densities and the current SFR density.  Finally we give our
summary and conclusions (Section~\ref{sec:summary}).

Throughout this paper we take $H_0 = 70 \,{\rm km\,s^{-1}\,Mpc^{-1}}$,
$\Omega_{m_0}=0.3$ and $\Omega_{\Lambda_0}=0.7$ for our cosmological
quantities, and where appropriate, define $h = H_0 / 100\,{\rm
  km\,s^{-1}\,Mpc^{-1}}$.


\section{Samples used in this analysis} 
\label{sec:data}

The Sloan Digital Sky Survey is a digital CCD survey in 5 optical
bands which intends to cover up to 10\,000 deg$^2$. An overview is
given by \cite{SDSS}. The imaging camera is described by
\cite{SDSSGunn}, the $ugriz$ photometric system and calibration by
\cite{SDSSFuk}, \cite{SDSSasinh}, \cite{SDSSHogg} and
\cite{SDSSSmith}. A large fraction of SDSS data is currently available
to the entire astronomical community \citep{SDSSedr}. The coordinate system
is defined to a precision of better than 0.1 arcsec \citep{SDSSAstrom}. The main galaxy
sample is essentially a magnitude limited spectroscopic sample.  It is
selected as virtually all galaxies in the photometric area with a Petrosian
magnitude $r<17.77$. The overall targeting completeness is 92\% \citep{SDSSTiling}. 
For galaxies in this magnitude range the Petrosian magnitude is close to
total, they are also close to `model magnitudes' (which are obtained
by profile fitting) which represent an alternative method of trying to 
estimate total magnitudes. Ninety-eight percent of
the galaxies span a redshift range of $0<z<0.25$ with a median
redshift of 0.10. Full details of the spectroscopic main galaxy sample
are given by \cite{sdssMGS}. The sample we consider in this paper uses
a total of 153\,000 galaxies selected to $r<17.70$ (magnitudes are
from the PHOTO v5.2 software corrected for Milky
Way reddening; see \cite{SDSSedr} for details) from the SDSS
spectroscopic data. 
The sample is data from 463 survey quality plates
in the North Galactic Cap which were observed between MJD 51433 and
5235 and 86\,000 galaxies lie in the range $0.01<z<0.11$, the principal
range considered in this paper. Only galaxies with secure redshifts
were selected.

The spectra are taken through 3.0-arcsec diameter fibers and have a
wavelength range of 3800-9200\AA\ and a spectral resolution
$\lambda/\Delta\lambda = 1800$ which is approximately constant across
the spectrum. The signal-to-noise ratio is $>4$ per pixel (pixels
width $\simeq$ 1--2 \AA). We use `smear' corrected spectra 
reduced using SPECTRO2D software v4.9. The smear procedure uses
dithered telescope pointings to make a low-order correction to
allow for aperture effects, it changes the large scale spectral
response at the $<\pm 2$\% level in the mean colors.

Following BG02, the spectra are combined in redshift slices (using the 
redshifts from the Princeton 1D spectra pipeline, \cite{Schlegel03}) by
scaling them to their $r$ luminosity and summing them. This scaling
provides for a first-order aperture correction. Thus in each slice
we get a total luminosity spectrum for all galaxies down to the 
luminosity limit of the survey at that redshift. Because of the
good spectrophotometry of SDSS it is not necessary to apply a correction
as was done by BG02.  

There are two important issues in using the cosmic spectrum to
constrain star-formation histories, the first is luminosity bias and
the second is aperture bias. These redshift surveys represent the
spectra of all galaxies down to a fixed {\em apparent} magnitude.  For
low redshift galaxies the magnitude limit will correspond to a faint luminosity.
Since luminosity densities from typical galaxy luminosity functions
(approximately \citeauthor{Schech76} (\citeyear{Schech76}) functions)
tend to converge for $M>M_*$, where $M_*$ is the Schechter `break'
absolute magnitude; once we have sampled more than 2--3 mags below
$M_*$ the observed cosmic spectrum would be very close to the total
spectrum. To be quantitative,
for a faint-end slope in the range $-1.3<\alpha < -1.0$
\citep{blanton01,Cole01,Cross02,Madgewick02}, integrating down to
$M_*+3$ shows that between 84--94\% of the light has been sampled. Thus we
expect samples to be highly complete in {\em luminosity} at low
redshift and give a similar result whether they are selected in $r$
or \bj.

\begin{figure*}
\centerline{\epsscale{1.5}\plotone{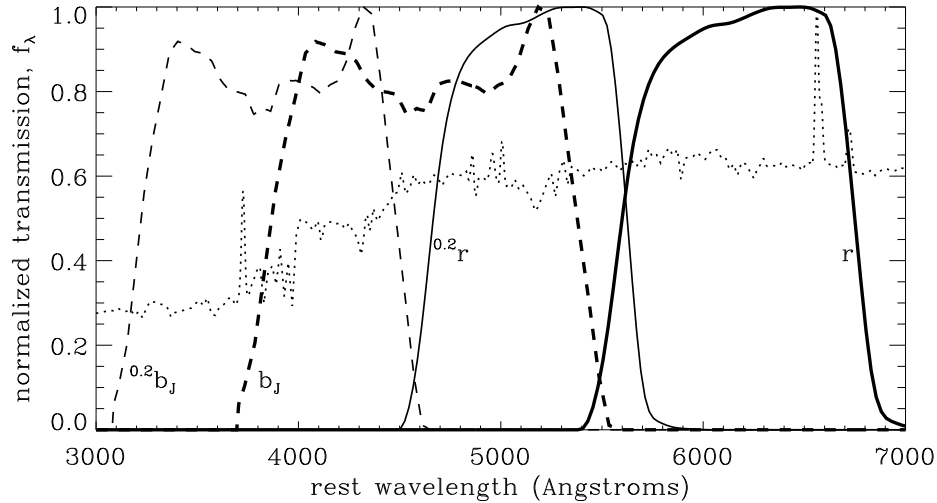}}
\begin{center}
\caption{Comparison of the selection functions of the two surveys. 
  The SDSS $r$ selection filter (solid lines, \citeauthor{SDSSedr})
  and 2dFGRS \bj\ selection filter (dashed lines, Hewett \& Warren
  private communication, these include atmosphere and telescope) 
  are shown at redshifts 0.0 and 0.2
  (bracketing the redshift range used in this paper).  At high
  redshifts, the \bj\ filter penetrates below the 4000\AA\ break and
  is therefore more sensitive to younger stellar populations. The
  dotted line is a theoretical cosmic spectrum from our analysis.  The
  $b_J$ curve was converted to photon-response equivalent for
  comparison with the $r$ curve.}
\label{fig:selection}
\end{center}
\end{figure*}

At high redshift we expect the blue/red sample selection to become
most prominent in the luminosity bias. In particular for the
redshift range we consider the SDSS red
selection does not sample below the 4000\AA\ break
(Figure~\ref{fig:selection}) where the UV from young stellar
populations dominates. Thus we would 
expect the relative bias to run
in favor of older SFHs in the SDSS sample as we approach $z \sim 0.1$
slices.

\begin{figure*}
\centerline{\plotone{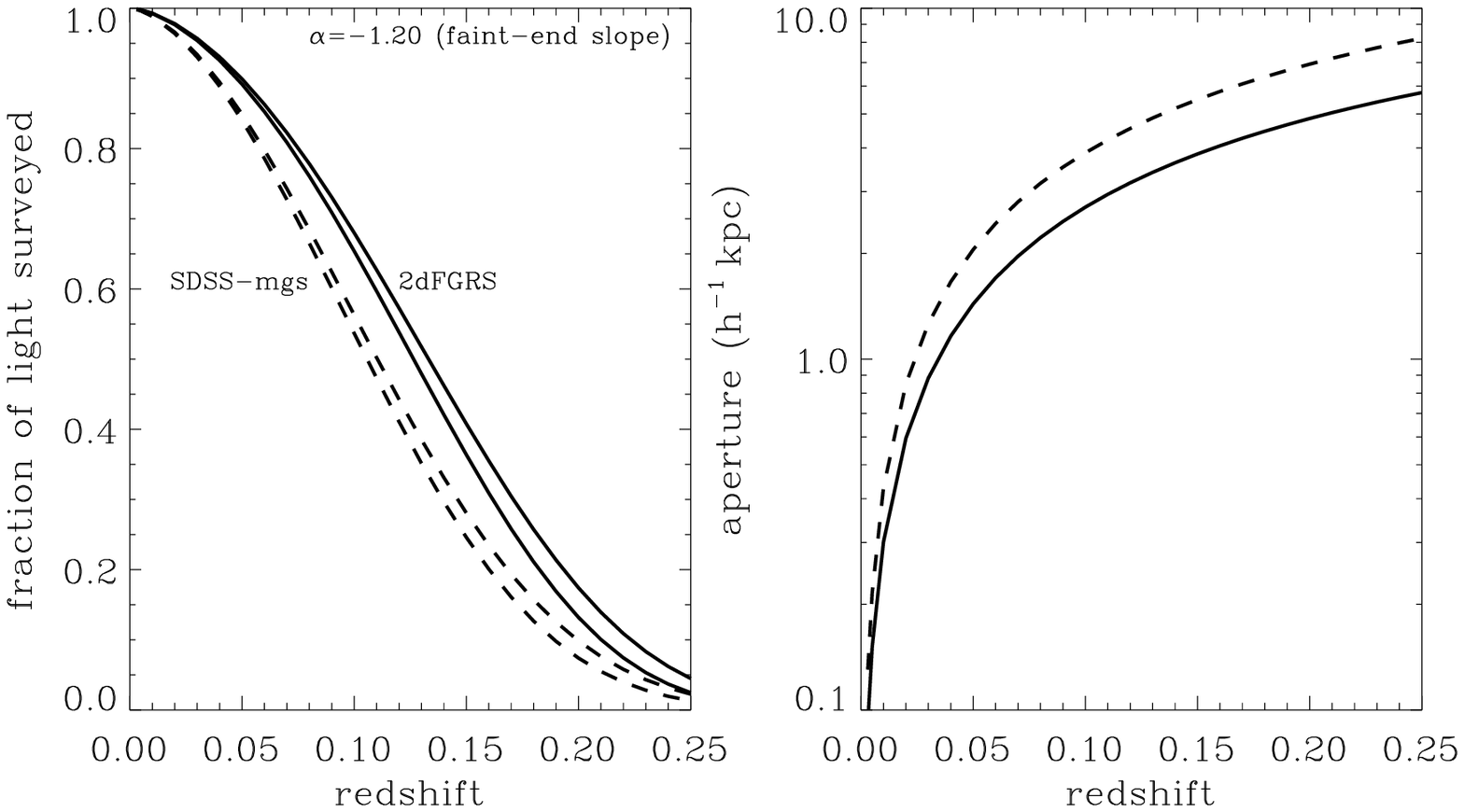}}
\caption{Comparison of sample bias in the SDSS and 2dFGRS redshift surveys.{}
  {\bf Left:} Luminosity bias versus redshift. This shows the fraction
  of total light of the galaxy population sampled down to the survey
  magnitude limit, calculated by integrating the appropriate
  \citeauthor{Schech76} luminosity function for each survey
  \citep[][both are consistent with a faint-end slope of
  $-1.2$]{blanton01,Madgewick02}. {\bf Right:} Projected aperture size
  versus redshift.}
\label{fig:sample}
\end{figure*}

In contrast, at low redshift we expect {\em aperture effects} to be
most apparent since the fixed size in arcseconds of the spectroscopic
fiber aperture corresponds to smaller physical scales in the galaxy.
At high redshift the effects are reversed --- the fiber apertures
ought to sample most of the light of a galaxy, but we are only seeing the
most luminous galaxies and red versus blue selection becomes more
important.  These sample biases are quantified in
Figure~\ref{fig:sample} which compares the two surveys.

The SDSS data gives us an opportunity to test for aperture bias in a way which was
not possible for the 2dFGRS work: because of the excellent SDSS
imaging and photometry we can use galaxy colors as a function of
aperture size as a diagnostic of aperture bias.

For comparison between the two samples at a common redshift, we define
the redshift limits so that the surveys penetrate to the same relative
depth below $M_*$. For example in the low redshift volume we use
$0.02\la z \la0.05$ for SDSS and $0.03\la z \la0.06$ for 2dFGRS, the redshift
upper limit corresponding to $M_*+2.6$ in both cases. The 2dFGRS
cosmic spectra are taken from those published by BG02. In
Table~\ref{tab:redshift-ranges} we define low, medium and high
redshift ranges (A, B and C) for which SDSS and 2dFGRS are
approximately equivalent in luminosity range.  Because the 2dFGRS
sample is slightly deeper than the SDSS sample, this correction also
works in the direction of minimizing the physical aperture difference,
since SDSS fibers are bigger (in arcseconds) than 2dFGRS fibers. This procedure
reduces the difference in physical aperture diameter in ranges A,B,C
from $\sim$50\% to $\sim$25\%.

\begin{table*}
\caption{Luminosity-bias equivalent redshift ranges (low/medium/high) 
  between the SDSS \& 2dFGRS samples}.
\label{tab:redshift-ranges}
\begin{center}
\begin{tabular}{c|c|c|c}
\hline
Region  &  2dFGRS redshift range  & SDSS redshift range 
& limiting magnitudes$^a$ \\
\hline
\hline
A & $0.025  <z\le 0.06$     & $0.015  <z\le 0.05$  & $M_* +4.5$ to $+2.6$ \\
B & $0.06~\,<z\le 0.10$     & $0.05~\,<z\le 0.08$  & $M_* +2.6$ to $+1.4$ \\
C & $0.10~\,<z\le 0.13$     & $0.08~\,<z\le 0.11$  & $M_* +1.4$ to $+0.7$ \\
\hline
\end{tabular}
\end{center} 
\hspace{1.2in}\vbox{\hsize = 5in {$^a$ Approximate range of limiting
    absolute magnitudes from the low-redshift cut to the high redshift
    cut relative to the Schechter-break luminosity}}.
\end{table*}

\begin{figure*}
\centerline{\epsscale{2}\plotone{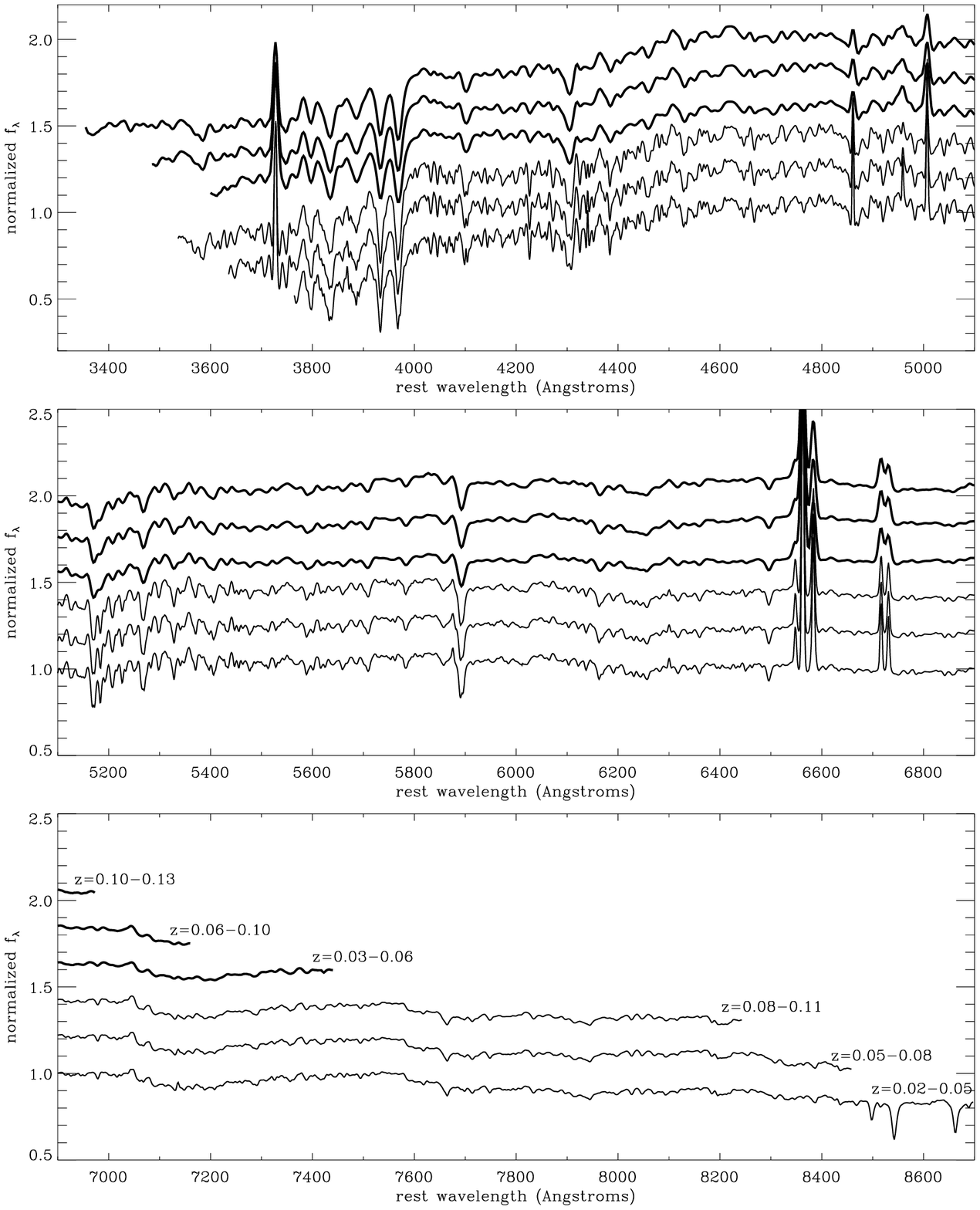}}\epsscale{1}
\caption{Comparison of SDSS and 2dFGRS cosmic spectra. 
  Three spectra are shown for regions A, B and C as defined in
  Table~\ref{tab:redshift-ranges}. The lower set (thin lines) are for
  SDSS and the upper set (thick lines) are for 2dFGRS (spectro-photometrically
  corrected). All are
  normalized by setting the mean flux between 4200 and 5800\AA\ to
  unity but are offset by multiples of 0.2 for clarity. Note the
  resolution of SDSS is two to three times better than 2dFGRS.}
\label{fig:spectra}
\end{figure*}

Figure~\ref{fig:spectra} shows a comparison between the normalized
cosmic spectra for SDSS and 2dFGRS for the volumes A,B,C.  The 2dFGRS
cosmic spectrum has been spectrophotometrically corrected against the
model fits as described by BG02.  The SDSS spectrum uses the native
spectrophotometry. Despite the differences in sample selection and
spectral resolution the cosmic spectra look very similar, especially
in the absorption features which we use to constrain the SFH.  To be
quantitative, in region A where we expect the selection bias to be
smallest, the RMS of the ratio of the two spectra over the wavelength
range 4000\AA--8000\AA\ is 2.6\% (comparable to the uncertainties in
the spectrophotometric modeling quoted by BG02), after smoothing the
SDSS spectrum to 2dFGRS resolution. We can break this down into
low-pass (i.e.\ continuum) and high-pass (line) variations using a
200\AA\ smoothing filter. For the smoothed spectra the RMS ratio is
2.4\% whereas in the spectra with the smooth component divided out the
RMS ratio is only 0.7\%. They key point is the high-pass stellar
absorption line information is almost identical in the two
spectra.\footnote{Another very low resolution comparison is provided by
  the \cite{CIE86} chromaticity values of the spectrum, we get (0.340,
  0.340) very close to the BG02 value of (0.345,0.345) and still close
  to white}  The similarity of the cosmic spectra from the two
surveys indicate the selection difference is minimal at low redshift.
Comparing with the uncertainties quoted by BG02 (which are mostly
systematic) we would expect our SFHs from SDSS to agree at the 95\%
confidence level with those of BG02.  It is this detailed modeling of
the SFH, to which we now turn.


\section{Star Formation Histories derived from cosmic spectra} 
\label{sec:results}

\subsection{Methods}

For our modeling of star-formation history we use the empirical
`double power-law' parameterization used by BG02. The
star-formation rate (SFR) is a simple function of redshift with a
break at redshift unity:
$$ {\rm SFR} \propto
\begin{array}{cl } 
(1+z)^\beta   &   \hbox{  for } z<1 \\
(1+z)^\alpha  &   \hbox{  for } 1\le z<5 \\
0             &   \hbox{  for } z\ge 5 \\
\end{array}
$$ 
The two power-laws are normalized at $z=1$ and star-formation is
started at $z=5$. This simple fitting form is chosen because it
already provides a good match to the range of observations of cosmic
SFH (for example, see figure~9 of \citeauthor{steidel99}
\citeyear{steidel99}), has a small number of parameters and provides
acceptable fits. Other parameterizations are possible, see BG02 for
some examples. All are principally measuring the ratio of old to young
stars weighted in some fashion.

The fitting of star-formation histories to the cosmic spectra proceeds
using standard evolutionary synthesis techniques and follows that of
BG02 for the 2dFGRS data. We use the PEGASE.2 evolutionary synthesis
models \citep{FR97}. We assume a universal Initial Mass Function (IMF)
which is independent of cosmic epoch and for this we use the IMF of
\cite{Salp55}. Evolutionary tracks are from the `Padova group'
\citep{bressan93} and the stellar atlas comes principally from the
theoretical stellar spectra of R.~L.\ Kurucz as given by \cite{LCB97}.
Metallicity evolution in this code follows the prescription of
\cite{WW95}. The fiducial extinction was chosen as that for a
inclination-averaged disk geometry but this makes little difference
because, as we will see below, our analysis primarily uses the high-pass
spectral information and we varied the extinction to test the effect
on the broadband color information.

We use an approach of consistent chemical evolution where the
interstellar medium for forming new stars is continuously enriched by
the death of old stars.  Using the PEGASE.2 closed box model, we
control this by a SFR normalization $r$ defined by BG02.  This
ranges from: $r\sim0.3$ where only a fraction of the available gas is
used to form stars so that the metallicity remains low, to; $r\sim1.4$
where a total mass of stars is formed over time greater than the mass
of gas {\em initially} available so that chemical evolution is
significantly faster.  The metallicity monotonically increases with
time and different values of $r$ correspond approximately to different
end-point metallicities. We quote end-point metallicity values from
PEGASE.2 which are luminosity weighted.  We implicitly assume that the
range of metallicities at each galaxy epoch can be represented by an
average metallicity for the purposes of evolutionary synthesis.  This
approach is more complicated than using simple-stellar populations
(constant metallicity) but is consistent with the logical assumption
that older stellar populations should, on average, have a lower
metallicity than younger populations.

\begin{figure*}
\centerline{\epsscale{0.8}
            \plotone{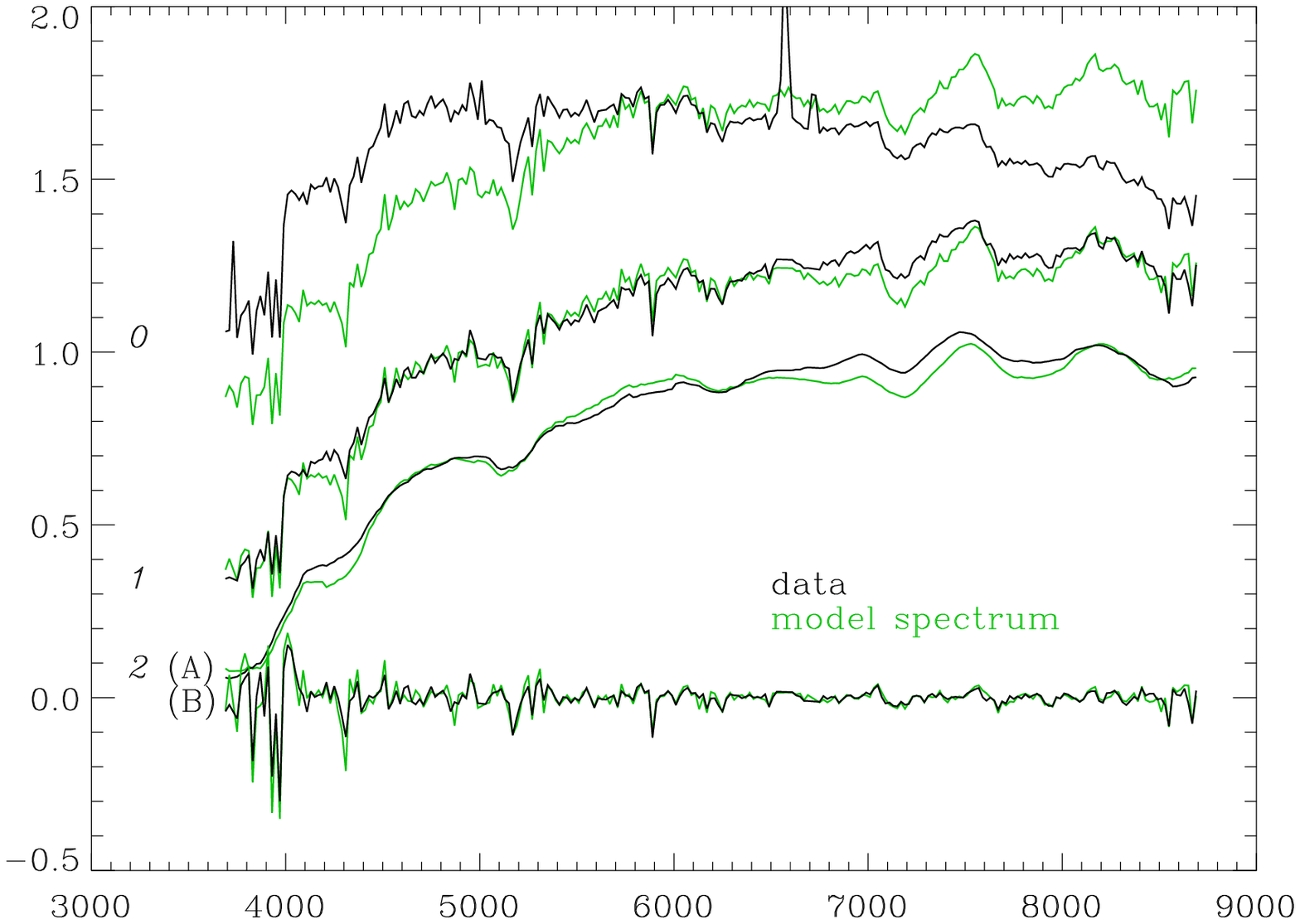}
            \plotone{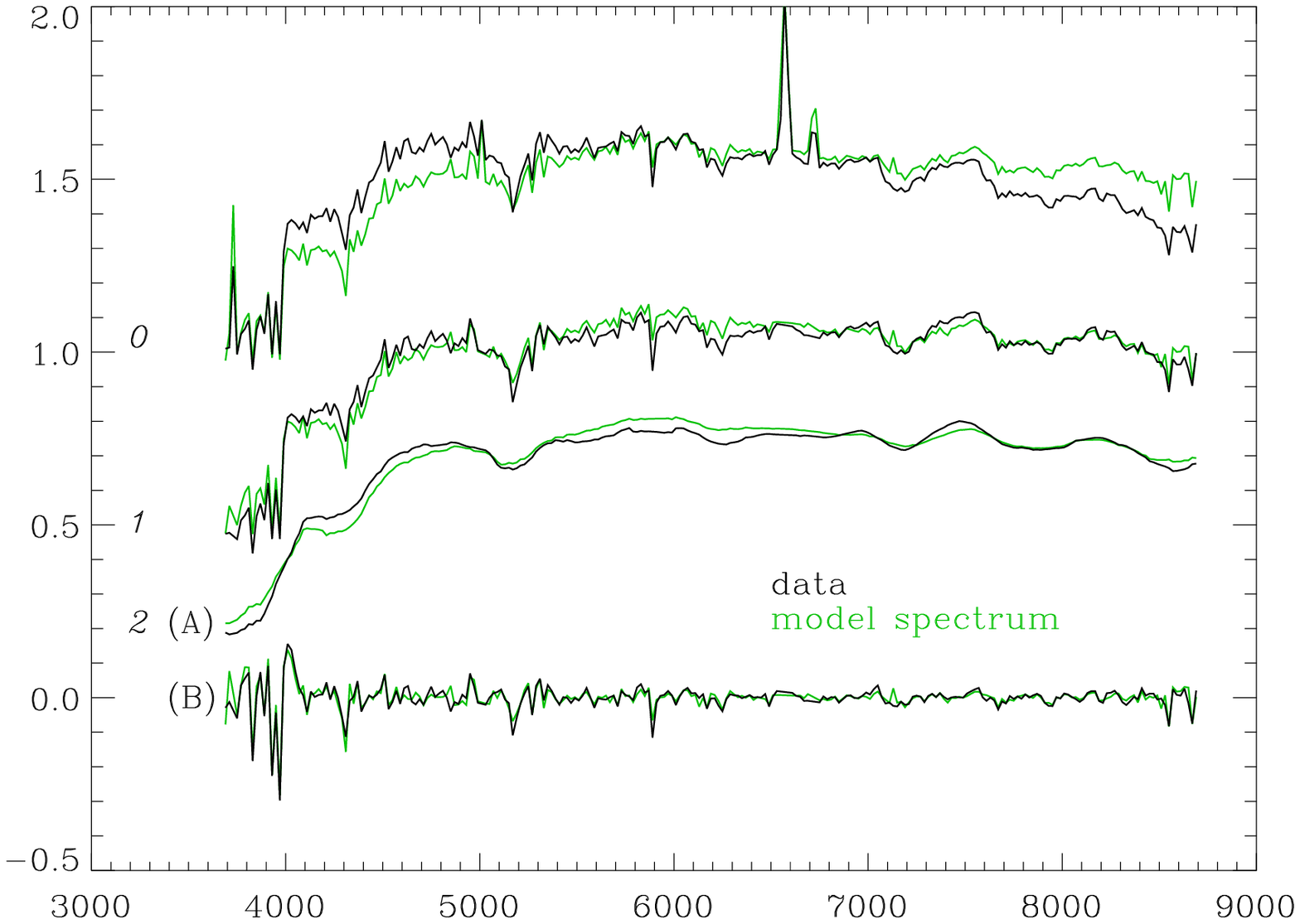}}
\centerline{\plotone{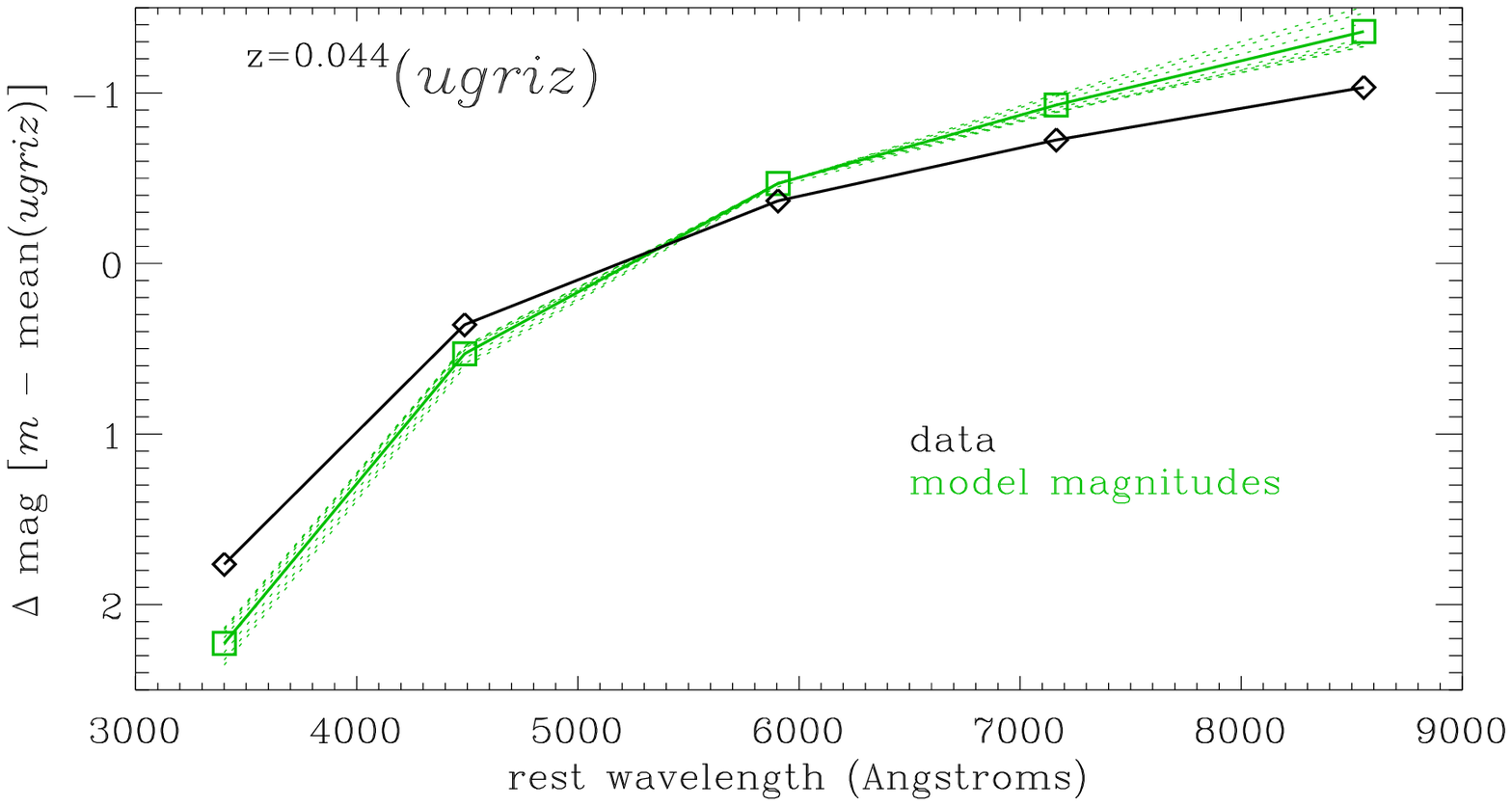}
            \plotone{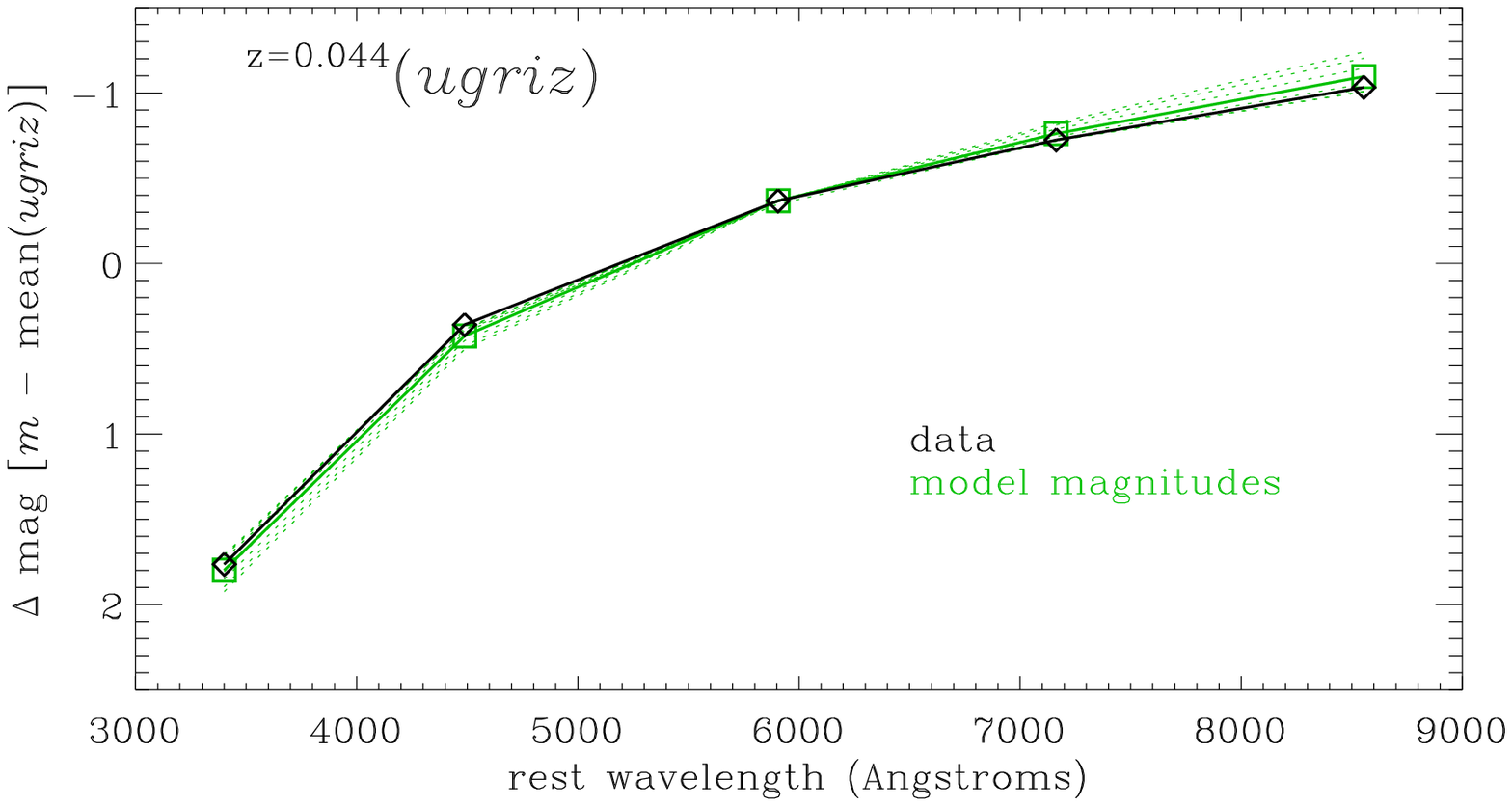}}
	    \epsscale{1}
\begin{center}
\caption{Determination of figures of merit (FOMs).
  (left column is a poor fit model, right column is a good fit model).
  {\bf Upper}: spectral FOMs. The normalized data and and example
  model spectrum (step {\it 0}, offset by $+0.5$) are adjusted by a
  low-order polynomial to remove power on broad-band scales and some
  emission lines are cut out to remove nebular components. This forms
  an adjusted set of spectra (step {\it 1}). To compare these, the
  spectra are further divided into low-pass and high-pass components
  (step {\it 2} A and B, offset by $-0.3$ and $-1.0$). The FOMs are
  determined from the $\chi^2$ difference between the model and data
  for each component.  {\bf Lower}: photometric FOM. The broad-band
  fluxes are coadded without $k$-corrections to form the data
  $^z(ugriz)$ magnitudes where $z$ is the mean redshift of the slice.
  The model magnitudes are calculated by integrating the model
  spectrum through the redshifted bandpasses.  Both are normalized by
  subtracting the mean of $ugriz$.  The FOM is formed from the minimum $\chi^2$
  difference after comparing the data with 10 different extinction
  versions of the model magnitudes (dotted lines). The squares
  represent the fiducial extinction model.}
\label{fig:analysis-methods}
\end{center}
\end{figure*}

One approach to fitting observed spectra to models would be to
compute line indices for both data and models (e.g. \cite{kauffmann02}).
The approach we use here is to fit the entire spectrum using a $\chi^2$
statistic.
For establishing our goodness of fit, we follow BG02 and compute
different $\chi^2$ figure of merits (FOMs) from high-pass and low-pass
filtered data. As in BG02 we calculate the figure-of-merit quantities by
summing over all wavelengths except near strong nebular emission lines
(O{\small II} 3727\AA; O{\small III} 5007\AA; H$\alpha$ 6563\AA; N{\small II}
6583\AA; S{\small II} 6716\AA\ \& 6730\AA) as we are only interested in the
stellar emission. The first step is to fit the spectrum with a 
2$^{\rm nd}$ order polynomial and divide the spectrum by this fit.
This effectively removes power on broad-band scales (i.e. $\sim$ 2000\AA\
resolution).
The spectra are then convolved with a 200\AA\ top-hat filter
to make a smoothed version and are then divided by this
smoothed version. Thus only high-pass spectral information is
retained. The $\chi^2$ is then the fit between the high-pass model and
the high-pass data, using a suitable noise model.  This is `FOM~B' in
the nomenclature of BG02.  Another $\chi^2$ is computed from the
low-pass  200\AA\ resolution spectra (FOM~A). Finally 
we also compute a $\chi^2$ fit between SDSS $ugriz$ colors and PEGASE
model colors which is illustrated in the lower panel of
Figure~\ref{fig:analysis-methods}. This final FOM is marginalized over a
range of extinction corresponding to about $A_V=0.5$--2 mags because
of the extinction dependency of this FOM.
The upper panel of
Figure~\ref{fig:analysis-methods} illustrates the steps in this
procedure for data and example model spectra (both bad fits and good
fits).

\begin{figure*}
\centerline{\epsscale{1.5}\plotone{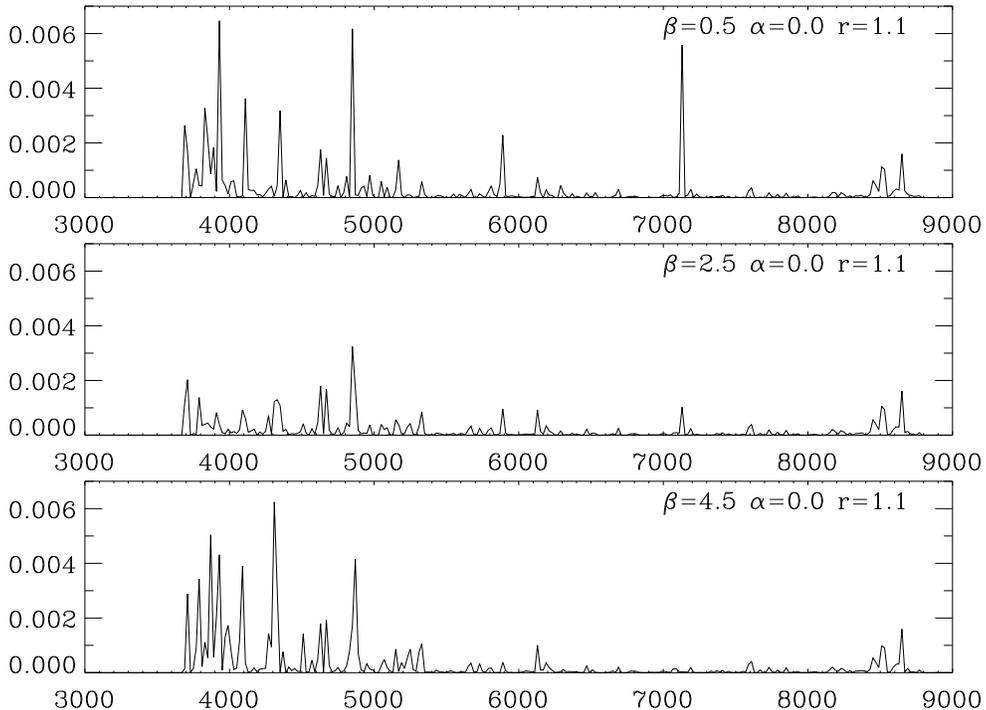}\epsscale{1}}
\caption{Example of residuals from the high-pass analysis: 
  residual squared versus wavelength. The middle panel represents a
  good-fitting model while the outliers show models rejected at very
  high formal confidence ($>99.99$\%) in comparison with the spectrum
  from SDSS region~A.}
\label{fig:residuals}
\end{figure*}

In general, we use only the high-pass FOM or a weighted combination of
the FOMs with less weight being given to the low-pass spectral /
broadband color information.  Examples of residuals for the high-pass
FOM are shown in Figure~\ref{fig:residuals}.  The advantage of this
approach is that low-pass systematic uncertainties such as flux
calibration errors and dust extinction are effectively excluded. The
disadvantage is that broadband color information is also mostly
excluded from the analysis. In practice we find this makes little
difference since most of the information in the colors is already
encoded in the spectral breaks and line indices. Additionly, we note
that regions near strong emission lines are excluded from the analysis
as we are interested in star-formation {\em history}, not the
instantaneous SFR that the emission lines are most sensitive to.

To estimate the errors, we follow the approach of BG02 and
divide the sample up into 10 contiguous sky areas. We then use the
variance between cosmic spectra from different sky areas to estimate
the errors, this has the advantage of including systematic effects.
Simply taking the raw variance will overestimate the true errors,
whilst dividing it by 10 will not properly take into account
systematic effects.  As a compromise, we estimate the confidence
limits with Monte Carlo simulations each time drawing 5 random entries
from the 10 regions.  This will in principle still overestimate the
errors by $\sqrt{2}$ but is robust against systematic effects. In
addition, we add $\sim 2$\% random errors to the models (with power on
all scales from high-pass spectra to broad-band photometry).

\subsection{Results}

BG02 found the 2dFGRS cosmic spectra exhibited no unique
solution but there was a broad curved degeneracy surface in the
$\alpha-\beta$ plane. However it did allow the defining of a common
solution where all the different measurements of $\alpha-\beta$
agreed.  As discussed by \cite{Hogg02} the surveys to $z=1$ favor
$\beta\simeq 3$. Direct UV observations of high redshift galaxies give
$\alpha \simeq -1$ \citep{madau96}, i.e.\ a rapid decline, but once
these are extinction corrected for dust, estimates range from constant
SFR with $\alpha=0$ \citep{madau96,steidel99} to increasing star formation, 
with $\alpha=1$ for $1<z<1.5$ \citep{CSB99}. 
The problem of the effect of dust extinction, and the sample bias
effects such as luminosity and surface brightness selection, is that
increasing amounts of UV are missed as one goes to higher redshifts.
It has recently been claimed that very large amounts of UV and hence
star formation are missed at high redshifts \citep{Lanzetta02}.  So
the values of $\alpha$ are really lower limits. This is where the
cosmic spectrum constraints become the most useful: only a narrow
range is allowed and high $\alpha-\beta$ is excluded. The conclusion
of BG02 was that there was a concordance SFH of
$1.8<\beta<2.9$ and $-1.0<\alpha<0.7$ that agreed with the various
determinations at the 95\% level.

\begin{figure*}
 \centerline{\plotone{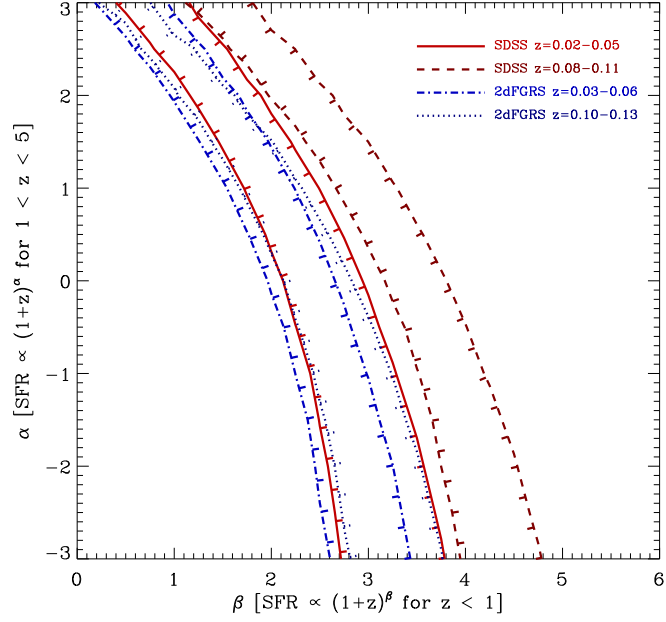}}
\caption{Comparison of star-formation histories derived from the SDSS and
  2dFGRS samples in the $\alpha-\beta$ plane for cosmic spectra from
  redshift regions A and~C. The contours represent 90\% confidence in
  our analysis using only the normalized high-pass spectra. For a
  given $\alpha=0$, the difference between the SDSS and 2dFGRS results
  amounts to $\Delta\beta \sim 0.25$ for the low-redshift samples and
  $\Delta\beta \sim 1$ for the high redshift samples.}
\label{fig:sfh1}
\end{figure*}

\begin{figure*}
   \centerline{\epsscale{1.5}\plotone{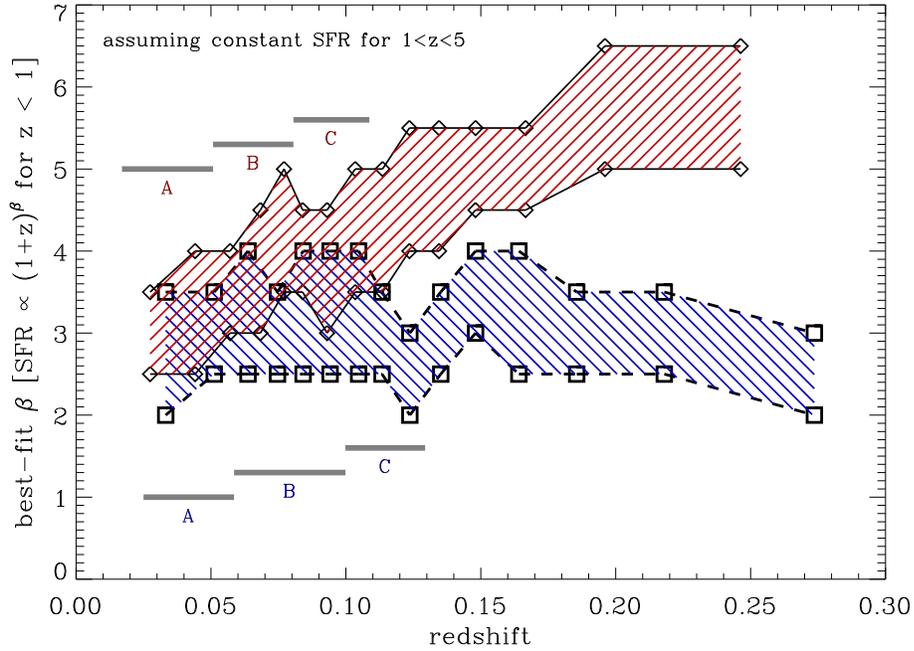}}
\caption{Comparison of star-formation histories derived from different 
  redshift slices.  For each slice we plot the 95\% confidence region
  on $\beta$, for $\alpha=0$. The upper region represents the SDSS
  contours (solid bounding lines) and the lower region represents the
  2dFGRS contours (dashed bounding lines). The variation in this plot
  does {\em NOT\/} represent any cosmic evolution, rather it shows
  primarily the effect of luminosity-selection bias in measuring
  cosmic SFH (represented by $\beta$). The divergence is due to the
  red versus blue selection in the samples as one goes out to high
  redshift.  At low redshift where most of the cosmic light is being
  counted by both surveys the values of recovered $\beta$ converge.
  The A,B,C lines show the redshift ranges defined in
  Table~\ref{tab:redshift-ranges}.  Note that the results here differ
  slightly from Fig.~\ref{fig:sfh1} because here we have marginalized
  over metallicity in the range 0.5-1.5 \Zsun.}
\label{fig:beta-z}
\end{figure*}

We compare SDSS and 2dFGRS derived cosmic SFHs by plotting both of
them in the $\alpha-\beta$ plane (with $r=1.1$).  This is shown in
Figure~\ref{fig:sfh1} for the two redshift ranges A and~C.  Contours
are shown at the 90\% confidence level with only the high-pass FOM
applied.  In all cases the general shapes of the allowed, degenerate
regions are similar, indicating qualitative agreement between the kind
of SFHs permitted by SDSS and 2dFGRS cosmic spectra. At low redshift
(region A), where we expect the samples to have the least luminosity
bias, the agreement is good and well within the formal 90\% confidence
limits. The difference amounts to $\Delta\beta \simeq 0.25$ for
$\alpha=0$ and fixed chemical evolution.  This leads to our first
important conclusion: the effects of sample selection in the 2dFGRS
paper were indeed small at low redshift.

If we look at the higher redshift contours (region C) then there is an
obvious shift in the contours from their low-redshift positions, which
is the effect of the aforementioned luminosity bias and sample
selection.  This becomes clearer if we can plot the trend with
redshift, to do this in Figure~\ref{fig:beta-z} we fix $\alpha=0$ and
plot $\beta$ versus redshift for 2dFGRS and SDSS redshift slices.  The
plot shows 95\% confidence regions.  Here, we apply a weighted
combination of low-pass/broadband and high-pass FOMs but qualitatively
the results are the same if only the high-pass information is used.
At low redshift, the two regions converge and overlap considerably
around $2<\beta<4$ whilst at high redshift the two diverge, which we
interpret as the effects of differential luminosity bias in the red versus blue
selection.  For the red sample the derived $\beta$ for a slice
increases considerably approaching $\beta\sim 6$ at $z\sim 0.2$,
indicating an increasing dominance of luminous but old red galaxies.
In the blue sample this is somewhat counteracted as the blue selection
below 4000\AA\ can include more young star-forming galaxies at these
redshifts leading to lower values of $\beta\sim 3$ at $z\sim 0.2$.
This can be seen directly in the colors: if we match up 2dFGRS
redshifts with SDSS photometry we also see much bluer $u-r$ colors for
a $z>0.2$ and \bj$<$19.45 sample (matching the 2dFGRS selection) than
for a $z>0.2$ and $r<17.7$ SDSS selection. The effect is strongest
for $z>0.2$ where the \bj\ band is pulled below the 4000\AA\ break.
We note that at low-redshift the $u-r$ distribution is bimodal
with a red-type and blue-type branch \citep{BAL03}
whereas at $z>0.2$ the blue peak is gone in the $r$ selected SDSS sample
but is still there in the \bj\ selected 2dFGRS sample.
The effect of producing a more constant $\beta$ is somewhat
fortuitous.  One can try making a blue-selected subset of the
SDSS spectra, however to be complete for all galaxy types it has to be
much brighter (approximately \bj$<$18.1) and the redshift distribution
is too low to show significant effects (98\% of galaxies have $z<0.15$
with a median redshift of 0.07).  However the important point is that
for the lowest redshift slices shown here we have a good approximation
to a complete volume limited sample and that the selection function
becomes small.  For $\alpha=0$ we find from the joint overlap of
$2.5<\beta<3.5$ (region A, 95\% confidence). If $\alpha$ is increased
then $\beta$ comes down.

We have demonstrated by comparison with the completely independent
SDSS sample that the effect of luminosity bias, one of the outstanding
issues in the 2dFGRS result of BG02, is small.  We note the
remaining problems of blue versus red samples could be improved,
particularly at high redshift, by a multiwavelength selected sample,
for example all galaxies {\em jointly} in each of $ugriz$ down to a
magnitude limit in each band which matched the estimated cosmic
spectrum at some redshift.

\begin{figure*}
\begin{center}
\centerline{\plotone{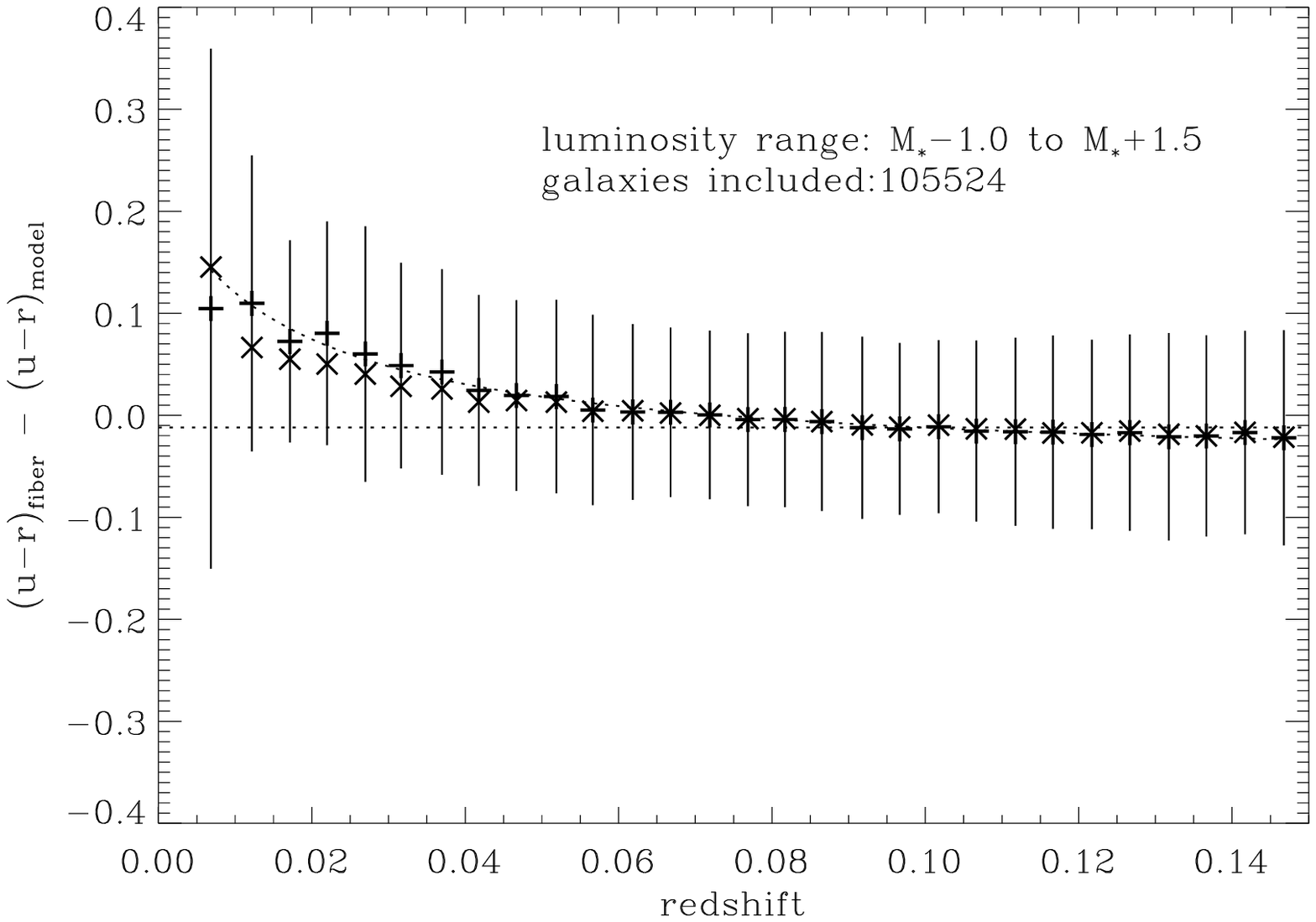}}
\caption{Plot of $(u-r)_{APER} - (u-r)_{MODEL}$ {\em average} colors
  for a fixed luminosity range in the SDSS sample versus redshift to
  illustrate aperture effects.  (See text for definition of our aperture
  magnitudes). The $\times$ symbols represent the
  median, the $+$ symbols the mean and the vertical lines denote the
  standard deviation (in redshift bins of 0.005). (We plot the latter instead
  of the standard error on the mean in order to illustrate the
  scatter). The median and mean
  values show a small offset at $z<0.05$ but it drops to zero at
  higher redshift. The dispersion remains large indicating for individual
  galaxies, as opposed to the mean, aperture effects remain a problem. We also
  note that the sample is not representative for $z>0.1$ (see text). }
\label{fig:aper-u-r}
\end{center}
\end{figure*}

The remaining important bias issue to consider is that of aperture
effects. The SDSS sample allows us to investigate this for the first
time using the colors. In Figure~\ref{fig:aper-u-r} we plot the
difference in the {\em average} $u-r$ color between `model' SDSS magnitudes and
`aperture' SDSS magnitudes which we have computed by integrating the
database image profile information to a 3 arcsec diameter aperture.
This approximates the SDSS `fiber' magnitudes but has the advantage of
being PSF corrected. We choose $u-r$ as it straddles the
4000\AA\ break and should be a color that is very sensitive to changes in
stellar populations \citep{strateva01}. The 3~arcsec diameter aperture matches the
spectroscopic aperture. The model magnitudes are determined by using
the best fit of de Vaucouleurs and exponential spatial profiles to
calculate a `total' magnitude. We do not use the SDSS Petrosian
magnitudes, although these work well for total magnitudes in
$griz$ bands we have determined there are significant problems
in the use of $u$ Petrosian magnitudes.\footnote{Specifically, the
  flux can be summed over large Petrosian radii causing significant
  magnitude errors because the $u$-band flux is so weak and objects
  are so close to the background level in the frames. This can be seen
  for example by comparing the distribution of $u-r$ with camera
  column. This is a problem with $u$ magnitudes from version 5.2 of
  the SDSS PHOTO software and may be improved in the next version.
  Please contact the authors for further information.} 

It can be seen from Figure~\ref{fig:aper-u-r} that there is no
significant difference in the {\em average} $u-r$ between model and fiber and no trend
with redshift for $z>0.05$. (We note however that at high redshift $z>0.1$ the SDSS
becomes dominated by early-type galaxies for which we naturally expect small
aperture effects in the colors). 
For $z<0.05$, there is a small aperture
effect amounting to $\Delta(u-r)\la 0.05$ (region~A).  We stress that for
{\em individual galaxies} there is a large dispersion in this statistic
because aperture bias is a significant problem for individual galaxies. Only
in the mean does the aperture bias appear  small. Indeed it appears
aperture bias from using this low-redshift sample is smaller than the
luminosity bias from using higher-redshift samples (region~B or~C).
Given there is no significant aperture effect in the colors, how does
this translate into stellar populations? As we know color is
degenerate with age and metallicity, so either the stellar populations
are the same or there is some fortuitous cancellation in age and
metallicity gradients.  The latter seems unlikely. The best evidence
is that aperture effects are not significant for the cosmic spectrum,
i.e.\ on average.  To investigate this in more detail 
would require two dimensional
resolved spectroscopy of a large number of galaxies.  We do note that
for individual galaxies, for example a face on spiral with a
significant old bulge population, aperture effects could be
significant. However, there are many types and different orientations
of galaxies in a magnitude-limited survey and the average fiber
placement on an average galaxy is more robust to aperture effects.

The final value of $\beta$ is still in the range of previous
estimates. From SDSS alone: for $\alpha=1$ we constrain
$2\la\beta\la3.5$, and; for $\alpha=0$ we constrain $2.5\la\beta\la4$.
These ranges cover metallicity degeneracy from about solar down to
half-solar with approximately 95\% confidence limits.  The $\beta$
values are close to the 2dFGRS results of BG02.  The low-redshift
values from the luminosity density measures in UV, nebular lines,
far-infrared and radio compiled by \cite{Hogg02} give a mean $\beta =
3.1\pm 0.7$ ($1\sigma$), with a spread in most of the measurements (5
out of 9) of $2.9 < \beta < 4.5$. The cosmic spectrum and luminosity
density measurements are entirely consistent, within their own
intrinsic uncertainties, and suggest that $\beta\sim 2$--4.

\subsection{Tests of high-redshift star formation}

\begin{figure*}
 \centerline{\plotone{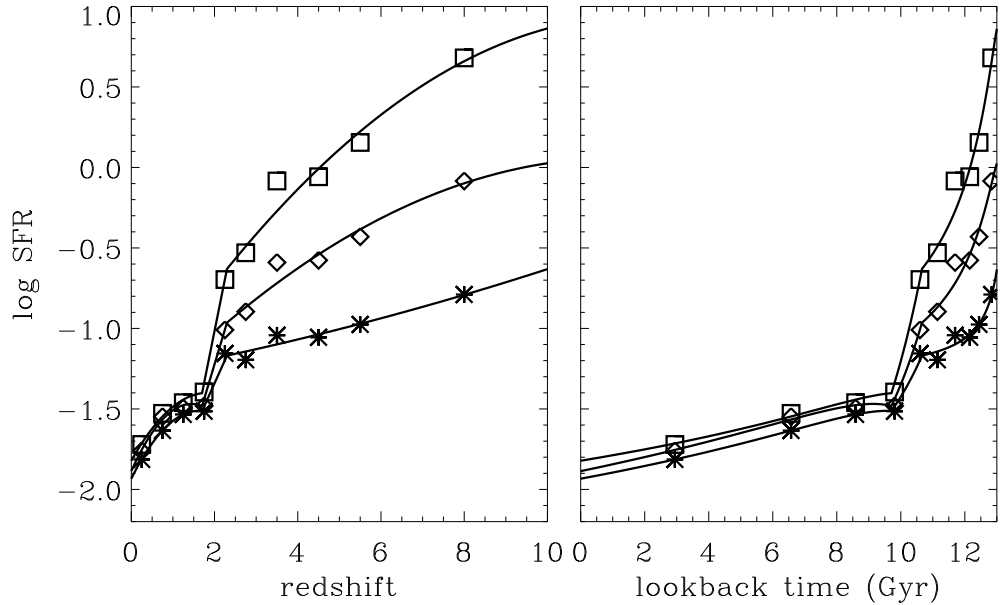}}
\caption{Three possible cosmic SFHs from the analysis of \cite{Lanzetta02}. 
  The symbols represent their data while the solid lines are our
  fitted to their SFH that we use in our test. The variation is
  principally at very early times, as illustrated in the right panel.
  The timescale shown is for the standard $\Lambda$ cosmology with
  $H_0=70$. Note that we only use the average values for each Lanzetta
  et al.\ model, i.e.\ we do not consider their error bars.}
\label{fig:lanzetta}
\end{figure*}

As well as the $\alpha-\beta$ parameterization we can test more
complex models of SFH. Recently \citep{Lanzetta02} proposed that
direct high-$z$ surveys were missing a large component of high
redshift light based on a comparison with the specific SFR intensity
distribution and the column density of Lyman-$\alpha$ absorbers which
they claimed were related. This is an ideal situation to test with the
cosmic spectrum because even if objects are obscured at high redshift
the stars produced must end up in objects in the low-redshift SDSS
data, unless of course they form some new local population yet to be
detected. We note that Lanzetta et al.\ proposed these {\em were} the
ancestors of todays galaxies.  They proposed three possible models
using different missing-light corrections, from different methods of
estimating the SFR intensity distribution function, which we term
`LOW', `MEDIUM' and `HIGH' based on their $z>2$ SFRs
(Figure~\ref{fig:lanzetta}). All have a low-z slope equivalent to
$\beta\approx1.2$. Note we do not use the $\alpha-\beta$
parameterization, rather we fit the SFHs given by Lanzetta's figure~4
directly with a break at $z\approx2$ (see Figure~\ref{fig:lanzetta}
where we replot the Lanzetta figure to illustrate the time dependence
and our fit to Lanzetta's points).  The results of fitting these SFHs
are given in Table~\ref{tab:lanz}. We use a weighted combination
of FOM~A with FOM~B in the ratio of  1-to-5 
based on uncertainties determined from the Monte-Carlo 
error estimation from our different sky regions. For reference we also give the
significance levels of a fiducial $\alpha=0$, $\beta=3$ model and a
$\alpha=1$, $\beta=2$ model.  The first gives a best-fit model with
about half-solar metallicity and the second with about solar
metallicity.

\begin{table*}
\caption{Comparison of SFH models with cosmic spectra}
\label{tab:lanz}
\begin{center}
\begin{tabular}{l|c|c} 
\hline
SFH Model  &  SDSS confidence$^a$  & 2dFGRS confidence$^a$ \\
\hline
\hline
$\alpha=0$, $\beta=3$  & \llap{$<$}68\rlap{.3}  &            68.3         \\
$\alpha=1$, $\beta=2$            & 80           &  \llap{$<$}68\rlap{.3}  \\
Lanzetta LOW                     & 98           &             68.3        \\
Lanzetta MED                     & 90            &            90           \\
Lanzetta HIGH                    & 99           &            99           \\
Lanzetta HIGH  $Z>0.5 Z_{\odot}$ & 99\rlap{.99} &            99\rlap{.99} \\
\hline
\end{tabular}
\end{center} \hspace{1.5in}\vbox{\hsize = 4in
  {$^a$ Percent rejection for models marginalized over metallicity.
    The models assume a universal Salpeter IMF and are compared with
    the low-redshift range cosmic spectra from the surveys. Lanzetta
    models are integrated for $z<10$.}}
\end{table*}

The HIGH model is rejected at $\ge 99$\% confidence for cosmic spectra
from both the SDSS and 2dFGRS surveys. In fact, the HIGH model only
fits at this level if the average metallicity becomes rather low
($Z<0.5 Z_{\odot}$), if we restrict the range to $Z>0.5 Z_{\odot}$,
the model is rejected very severely. The LOW model is also rejected by
the SDSS data because of the low $\beta$, though more marginally.
None of these models fit the cosmic spectra as well as our fiducial
$\alpha=0$, $\beta=3$ model which is well motivated by the high
redshift luminosity density measurements. Thus it appears the cosmic
spectrum does not provide any strong evidence for large amounts of
missing light in the current high redshift census. Quantitatively, no
more than about 85\% of stars formed at $z>1$. This is slightly higher
than the BG02 upper limit of 80\%. The differences are that we include
the SDSS data, include only the 2dFGRS low-redshift range, give some weight to
the low-pass components of the cosmic spectra and allow $\zform=10$
for the Lanzetta models. It is of course possible that some of our
assumptions may be wrong, such as a universal IMF; however the data
can be well fitted by models which assume a Universal IMF. The slope
of the high-mass IMF can be constrained by cosmic spectra, if near-IR
data is included. This will be addressed by a forthcoming paper
\citep{BG03}.


\section{The Absolute Cosmic Spectrum} 
\label{sec:phys}

We have shown that we have derived a normalized cosmic spectrum which
is consistent between SDSS and 2dFGRS surveys, is close to being
volume limited and is robust against aperture effects (by
comparing with SDSS $u-r$ colors).  An advantage of the SDSS
survey is the excellent spectrophotometry as standard stars are
included on each plate. In particular we would expect that the
relative fluxing with wavelength of the SDSS cosmic spectrum should be
much more accurate than for 2dFGRS (BG02 indicate 5--10\% errors are
likely in the latter). We do indeed find that the SDSS cosmic spectrum
is in agreement with the 2dFGRS cosmic spectrum {\em once a good fit
  spectrophotometric correction} has been applied to the latter (see
BG02 for details of this procedure, essentially it uses a good fit
high-pass model to set the fluxing).

\begin{figure*}
\centerline{\epsscale{1.}\plotone{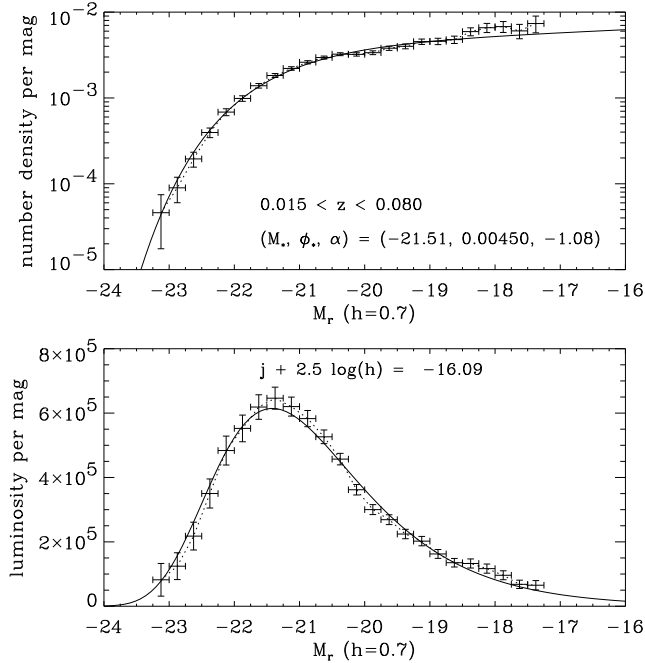}}
\caption{Luminosity function determined from redshift regions A+B.{} 
  {\bf Upper:} The number of galaxies Mpc$^{-3}$\,mag$^{-1}$ versus
  magnitude with a Schechter function fit (solid line). The vertical
  bars represent $\pm3\sigma$ Poisson errors. {\bf Lower:} A linear
  scaling of luminosity ($10^{-2.5/{\cal M}}$) versus magnitude, where
  ${\cal M}$ is the total absolute magnitude Mpc$^{-3}$\,mag$^{-1}$.
  This shows that the integration of the Schechter function is a good
  approximation to the luminosity density. This plot uses Petrosian
  magnitudes but the result is similar if Sersic magnitudes
  \citep{Blanton02} are used: $j+2.5\log h = -16.11$.}
\label{fig:lum-func}
\end{figure*}

So far we have been dealing with the normalized cosmic spectrum, and
principally fitting to the high-pass spectral information. However we
can put the cosmic spectrum on an absolute luminosity scale by
normalizing to the $r$-band luminosity density. For consistency, we
calculate this ourselves from a more recent large-scale structure
(LSS) sample by calculating the $r$ luminosity function using
$V_{\rm survey}/V_{\rm max}$ weighting and $k$-correcting to the
rest-frame $r$-band \citep[{\tt kcorrect v1\_11},][]{Blanton02-Kcorr}.
The LSS sample is similar to our cosmic-spectra sample but has a well
defined area, includes nearest-neighbor redshifts to replace galaxies
missed due to fiber collisions and has stricter limits on the
selection magnitude of $14.5<r<17.5$ \citep[taken from {\tt sample10}
described by][]{Blanton02}.  A Schechter function fits very well and
the total luminosity density is calculated from the analytic total
integration (Figure~\ref{fig:lum-func}).  We note there is a small
excess of galaxies above the Schechter function fit at low luminosities $M_r>-18.5$, a disrepancy
that has been noted by other authors \citep{Madgewick02} but
this has negligible effect on the final luminiosity density  
(see the lower panel of Figure~\ref{fig:lum-func}).
Regions A and B sample
better the faint and bright end of the luminosity function
respectively, the luminosity density in these two redshift ranges
agrees to $\pm 6$\%.  For our final $r$ luminosity density we take
the value calculated from regions A+B ($0.015<z<0.08$), this is $1.94
\pm 0.20 \times 10^8$ $h$ $L_{\odot}$ Mpc$^{-3}$ ($j+2.5\log h = -16.1
\pm 0.1$ mags where $j$ is the absolute magnitude of the integrated
light per Mpc$^3$).  
We estimate that the systematic uncertainties (redshift
ranges, Schechter fitting) are about 5--10\% and quote a 10\% error
bar. The final value agrees well with the more sophisticated
luminosity density calculation of \cite{Blanton02}.

We normalize our cosmic spectrum for region A (where it is the least
affected by luminosity bias) to this $r$ luminosity density by
integrating the flux of our cosmic spectrum through the $r$-band
filter profile \citep{SDSSGunn} and calculating a scaling factor. We
note that this will correct to first order for any residual luminosity
bias, as the luminosity density is calculated by integrating the
Schechter function fit to zero luminosity.

\begin{figure*}[ht]
\centerline{\epsscale{2}\plotone{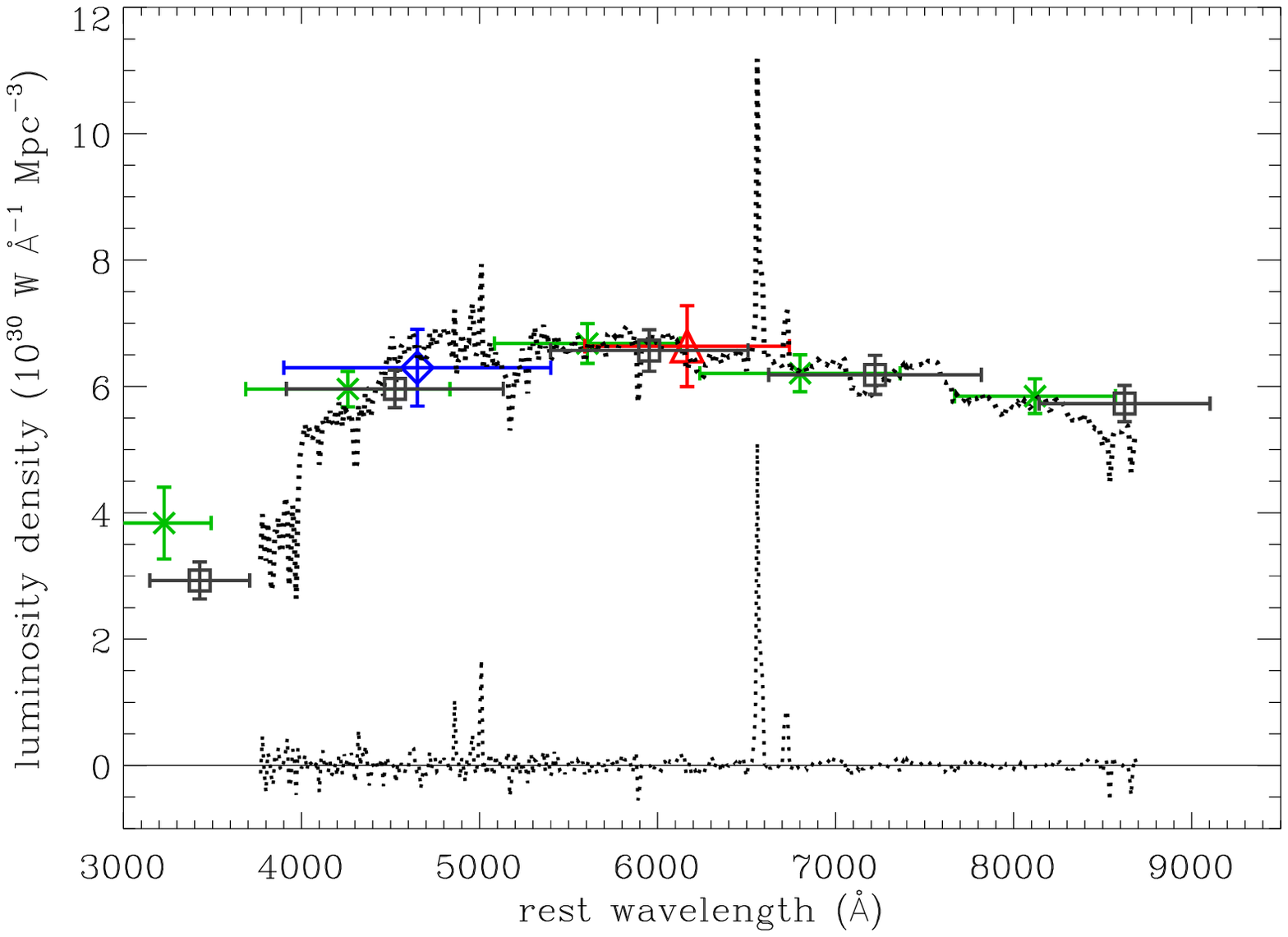}}
\caption{Absolute cosmic spectrum from SDSS region~A, i.e.{} 
  luminosity density per unit wavelength in physical units (dotted
  line, $h=0.7$).  This has been resampled at 10 Angstroms for display
  clarity.  The normalized cosmic spectrum (Fig.~\ref{fig:spectra})
  was scaled to match the $r$-band luminosity density
  (Fig.~\ref{fig:lum-func}) shown by the {\em triangle}. The {\em
    squares} show the relative luminosity densities, i.e.\ 
  flux-weighted observed colors, from region~A scaled to the same
  absolute luminosity density (note they are offset in wavelength to
  account for the mean redshift of 0.035).  The other points represent
  different estimates of broad-band luminosity densities: {\em
    asterisks} show the redshifted $^{0.1}(ugriz)$ filter 
    results of \cite{Blanton02}
  and the {\em diamond} shows the \bj\ point of \cite{norberg02}. The
  lower spectrum shows the stellar-component-subtracted cosmic
  spectrum described in the text (shown at the same resolution as
  above but line luminosities were estimated from the full resolution
  spectrum as in Fig.~\ref{fig:spectra}). This is still in physical
  units for the line fluxes, i.e. the ordinate units still applies.}
\label{fig:abs-spec}
\end{figure*}

This absolute cosmic spectrum (units: Watts \AA$^{-1}$ Mpc$^{-3}$) is
shown in Figure~\ref{fig:abs-spec} (smoothed slightly for the plot)
and tabulated in Tables A1 and A2.
We can check the reliability of this by comparing the spectrum with
luminosity densities computed in other broad bands, these should
correspond to smoothed estimates of the cosmic spectrum. In the
figure, we overlay the SDSS $ugriz$ luminosity densities
from \cite{Blanton02}\footnote{This is based on a sample of 80\,000
  galaxies and represents an update on \cite{blanton01}}, the \bj\ 
luminosity density from \cite{norberg02}, and the $ugriz$
luminosity densities calculated by using the cosmic-spectrum sample
directly (i.e.\ combining the photometric $ugriz$ fluxes
with the same weighting as the spectra).

There is excellent agreement within the errors. We note that the
luminosity densities given by \cite{blanton01} are 20--30\% discrepant
in $u$ and $g$, but the more recent determination shown here from
the \cite{Blanton02} sample of 150,000 galaxies are in much 
better agreement with this
work, with \cite{Cross02}, \cite{norberg02} and the revised SDSS work
of \cite{Yasuda}.  We note some minor differences in analysis: Blanton
et al.\ uses a maximum likelihood method which allows for evolutionary
effects (in a set of magnitude-limited samples
in each band separately) whereas the Cross \& Driver analysis used the
$1/V_{max}$ method, and the redshift ranges did not quite match
(although both were more or less $0<z<0.15$).

Finally we use our physical cosmic spectrum to derive some interesting
quantities. The mass/light ratio of our best-fit models is 3.7--7.5
$\Msun/\Lsun$ in the $r$-band (assuming $A_V \approx 1$). The range
corresponds to the range from a low-metallicity $\alpha=0$ model to
the solar metallicity MEDIUM Lanzetta model. Thus from the $r$-band
luminosity density we can derive the cosmological mass density in
stars: $\omstars h = 0.0025$--0.0055. Of course this is highly 
dependent  on the assumed Salpeter IMF as the galaxian light
is dominated by the most luminous stars. For example the mass-to-light ratio for an
IMF with slopes $(-0.5,-1.35)$ broken at 1 \Msun\  is about 60\% of the
unbroken Salpeter IMF $(-1.35)$. We also computed $\omstars h$ for the 
\cite{Kenn83} IMF $(-0.4,-1.5)$, redoing the fitting, and obtained 
0.001--0.002.

Similarly our best fitting range of models also gives us a mean cosmological SFR
today, the value range is 0.01--0.04 \sfrunits.  For a particular choice
of $\alpha$ and $\beta$ the range is narrowed as there is only a remaining
degeneracy with metallicity. For example if motivated by the high-redshift
luminosity density measures we choose $\alpha=0$, $\beta=3$ we find the
SFR today is 0.025--0.03 \sfrunits. It is also
interesting to look at the emission lines in the spectrum, by
subtracting the stellar `continuum'\footnote{By `continuum', we mean
  stellar continuum and absorption features.} and integrating the flux
we can derive the local luminosity densities of a whole set of useful
lines: [OII], H$\beta$, [OIII], H$\alpha$, [NII] and others.  We can
do a continuum subtraction by subtracting our best fit SFH stellar
model, although the resolution is different the subtraction is still
close to zero except near the nebular lines.  We used the best fit
high-pass model, so as to get the best match to absorption features,
and normalize it to the physical cosmic spectrum using a smoothing
filter.  We show the example continuum subtracted spectrum in
Figure~\ref{fig:abs-spec}.  The subtraction is excellent in the
4000\AA--8000\AA\ region, the principal discrepancy is for wavelengths
$>8000$\AA\ where the Calcium triplet absorption lines are not well
fit by any of our models; this effect can also be seen in 
Figure~\ref{fig:residuals}.

The line luminosity densities are simply
calculated by integrating the cosmic spectrum for each line from
$\lambda-8$\AA\ to $\lambda+8$\AA.  The box width is chosen to be
3$\times$ the typical line FWHM so the fluxes are very close to total.

\begin{table*}
\caption{Luminosity density in various lines derived from the cosmic spectrum}
\label{tab:lines}
\begin{center}
\begin{tabular}{lc|cc}
\hline
Line & & \multicolumn{2}{c}{Luminosity Density ($10^{30}$ W Mpc$^{-3}$)}\\
     & & ~~~Region~A~~~ & ~~~Region~B~~~ \\
\hline\hline
[OII]         & 3727\AA     & ---          &   $28 \pm 4$ \\
H$\beta$      & 4861\AA     &   $11 \pm 2$ & ~\,$9 \pm 1$ \\
\hbox{[OIII]} & 4959\AA     & ~\,$6 \pm 2$ & ~\,$3 \pm 2$ \\
\hbox{[OIII]} & 5007\AA     &   $22 \pm 2$ &   $13 \pm 2$ \\
\hbox{[OI]}   & 6300\AA     & ~\,$2 \pm 1$ & ~\,$2 \pm 1$ \\
\hbox{[NII]}  & 6548\AA     &   $11 \pm 1$ &   $11 \pm 1$ \\
H$\alpha$     & 6563\AA     &   $70 \pm 1$ &   $59 \pm 1$ \\
\hbox{[NII]}  & 6583\AA     &   $30 \pm 1$ &   $29 \pm 1$ \\
\hbox{[SII]}  & 6716\AA     &   $13 \pm 1$ &   $11 \pm 1$ \\
\hbox{[SII]}  & 6731\AA     &   $10 \pm 2$ & ~\,$8 \pm 2$  \\
\hline
\end{tabular}
\end{center}
\hspace{1.5in}\vbox{\hsize = 4in Notes: (i) Values are quoted for
  $h=0.7$. Errors on each value are for the continuum subtraction
  error; (ii) Additional errors of $\pm 10$\% should be added to allow
  for the systematic uncertainty in the $r$ luminosity density.}
\end{table*}

The resulting line luminosity densities are given in
Table~\ref{tab:lines}. We quote values for region A and B, as the
latter includes [OII] and for comparative purposes.  Both are
normalized to the same $r$ luminosity density.  There are two
sources of error: firstly the imperfection of the continuum
subtraction. Visually we checked the region around each line,
particularly the Balmer lines, and found no evidence of significant
residuals, i.e. the plot appears as an emission line with zero
continuum plus `noise' due
to the mismatch in resolution between raw data and model spectrum. 
To quantify this for each line we similarly extract two
regions, either side of the line, of the same width where there is
only blank continuum, i.e.\ typically a 40-60\AA\ offset. (These are
optimized for each line particularly in the crowded H$\alpha$/[NII]
region).  These give typically the same error and we quote the maximum
in the table. There is also an additional systematic error due to the
uncertainty in the $r$ luminosity density which we estimate as 10\%.

The line luminosity densities lead to some interesting results: first
the Balmer decrement H$\alpha$/H$\beta$ is $6.4\pm 0.9$ (region A but
region B is similar), if we take an unreddened case B recombination
value of 2.86 \citep{HummerStorey} and a Milky Way dust law from
\cite{Pei} (the SMC law gives similar numbers as the two are close in
the optical) we derive a nebular extinction $A_V=2.4\pm 0.4$. The
$H\alpha$ luminosity density thus requires a correction
factor of $\times (5.8\pm1.8)$. This is consistent with other
workers findings --- for example we have re-analyzed the sample
of \cite{Gallego95} looking at the $H\alpha$ correction factor as a function
of luminosity and find a volume averaged value of $\times 4.6$. Thus it
makes little difference whether one dereddens before taking the
mean or simply dereddens the mean (which is the approach we are
taking here with the cosmic spectum) even for these large nebular
extinction values.

The dereddened $H\alpha$ luminosity density is thus $4.1\pm 1.3
\times10^{32}$ W Mpc$^{-3}$ for $h=0.7$. Converting to cgs units and
$h=0.5$ this gives $2.9 \pm 0.9 \times10^{39}$ ergs s$^{-1}$ Mpc$^{-3}$
which is the same as that found by from the Canada France Redshift
Survey \citep{TresseMaddox98} at low redshift ($z\sim 0.2$).  
It is twice as high as that found by the
objective prism survey of \cite{Gallego95}.

From our models we can also work out the conversion of H$\alpha$
luminosity into SFR (from the number of ionizing photons), the mean
conversion factor is $1.36\times 10^{41}$ ergs s$^{-1}$ \Msun$^{-1}$
yr and the range in the models is $\pm 15$\% (it varies with
metallicity). This allows us to derive a SFR density today of $0.030$
-- $0.056$ \sfrunits.  It is remarkable to note that this is entirely
consistent with the derived independently from the star-formation
history fitting which illustrates the validity of our general
approach. The H$\alpha$ line independently measures the instantaneous SFR from young
ionizing stars, whereas the SFH fits a model to the stellar continuum
and is averaged over several Gyr. It is interesting to note that 
given the cosmic spectrum to get a SFR from SFH fitting much lower than 0.03 \sfrunits\
requires high $\beta$, high $\alpha$ AND a low metallicity today.

We note that the \cite{Gallego95} sample also gives a lower SFR
density than the UV sample of \cite{Treyer98}. Both \cite{Treyer98}
and \cite{TresseMaddox98} are measured at $z\sim 0.2$, whereas the
current sample is $z\sim 0.05$ which is comparable to the redshifts of
the \cite{Gallego95} sample. It is only the latter which is
discrepant. The most likely explanation is that the \cite{Gallego95}
survey covers only $1/5^{\rm th}$ our sky area and is only sensitive
to star-forming galaxies as it is emission line selected. Our survey
represents a `cosmic average' and galaxies contribute to the final
cosmic spectrum even if the H$\alpha$ flux is not detectable in
individual galaxies. We note that H$\alpha$ is only sensitive to
transient star-formation over $\sim 20$ Myr whereas a UV sample like
those of \cite{Treyer98} effectively average over longer times scales
of up to $\sim$ 1 Gyr (see \cite{glaze99} for a discussion of this).

A final comment on the other line ratios: the {\em observed\/}
H$\alpha/$[OII] line ratio of 2.1 (region B) is entirely consistent
with the median found by the sample of galaxies observed by
\cite{Kenn92b}. The other line ratios are also entirely consistent
with those from star-forming galaxies \citep{VeilleuxOsterbrock87}
including the weak [OI] line, indicating that the AGN contribution to
the cosmic spectrum is indeed negligible. It can at most be only
a few percent according to the models of \cite{kewley}. Most of the
optical light of the Universe does indeed come from stellar nucleosynthesis.


\section{Summary and Conclusions}
\label{sec:summary}


Computing the light in the Universe today from the SDSS and 2dFGRS
surveys allows us to derive a cosmic optical
spectrum, determine its robustness, and make more accurate
determinations of allowable star-formation histories of the Universe.
In particular:

\begin{enumerate}
\item We find a range of solutions $2<\beta<3$ and $0<\alpha<1$ which
  are consistent with the cosmic spectrum and determinations of the
  evolution of the luminosity density in various bands.
\item `Fossil Cosmology' and direct cosmology agree: i.e.\ the SFH
  inferred from the local Universe agrees with that measured by
  luminous emission at high redshift. Again the Copernican Principle, that 
  we are in no special place in the universe,
  is demonstrated by observational astronomy.
\item There is good agreement between SDSS and 2dFGRS derived cosmic
  spectra and SFHs at low-redshift, where we expect luminosity biases
  to be minimal, despite the difference in aperture. We conclude the
  result is robust against luminosity selection effects.
\item The excellent photometry of SDSS allows us to test for aperture
  effects; for average quantities such as the mean color and hence the cosmic
  spectrum we find the aperture effects are not significant (though they
  are still important for individual galaxies).  
\item Due to the excellent spectrophotometric quality of the SDSS data
  we can make an absolutely calibrated cosmic spectrum for a close to
  volume limited sample.
\item None of the SFH scenarios proposed by \cite{Lanzetta02} fit the
  cosmic spectrum as well as more standard models. There appears to be
  no compelling evidence for missing star-formation at high redshift
  from the cosmic spectrum.
\item Typically good fits to the cosmic spectrum (using consistent
  chemical evolution) give a final metallicity of 0.5--1 \Zsun.
\item We find the stellar population of the universe today has a best
  fit $r$-band mass/light ratio (for Salpeter IMF) of 3.7--7.5 $\Msun/\Lsun$, 
  which given the $r$-band luminosity density of $j+2.5\log h = -16.1$ gives
  $\omstars h = 0.0025$--0.0055 (and a factor of 2
  lower for a Kennicutt IMF).
\item By fitting the best stellar population model and subtracting, we
  can derive a whole set of nebular line luminosity densities for our
  cosmological volume. In particular we find the local dereddened
  H$\alpha$ luminosity density is twice as high as found by
  \cite{Gallego95} and similar to that found by \cite{TresseMaddox98}.
\item We find the SFR of the Universe today (for Salpeter IMF) is 0.03--0.04 \sfrunits,
  and agrees between models fit to the stellar population and that
  derived from the H$\alpha$ luminosity density of the same sample.
\end{enumerate} 

\subsection{Future Work}

One avenue which it is clearly possible to explore is how the SFH
varies with the luminosity of galaxies. The cosmic spectrum represents
a binned approach, i.e.\ looking at the ensemble stellar population in
a volume and inferring the SFH. We can not specify which galaxies the
stars were in at earlier times, i.e.\ it is not sensitive to merging.
We can extend this approach by computing the cosmic spectrum in
luminosity bins, i.e.\ estimating the SFH of stellar populations as a
function, approximately, of the mass of the galaxy they end up in
today. Does the SFH depend significantly on this? Given known
color-luminosity relationships which extend across a range of Hubble
types \citep[e.g.][figure~4]{Gavazzi93} obviously we expect it too.

In fact we have seen already some evidence for this in the current
paper, where the selection effect of the magnitude limit causes the
high redshift part of the SDSS survey to have higher luminosities and
steeper values of $\beta$. Luminous galaxies on average contain older
stellar populations. In the next paper in this series \citep{BAL03}
we present a detailed analysis of this differential
SFH in today's Universe.

One limitation of this work is the assumption of a universal Salpeter
IMF. The high-mass IMF can be constrained by cosmic spectra, if near-IR
data is included. This will be addressed by a forthcoming paper
(Baldry et al., 2003).

Another limitation of the current work is the dependence on relatively low
resolution (20\AA) models of evolutionary synthesis, whereas the SDSS
data has 2--5\AA\ resolution. The SDSS cosmic spectrum resolves many
low equivalent width metal lines which are not exploited by the
current low resolution analysis and will help to break the
age-metallicity degeneracy.  We are working toward being able to
construct high-resolution models to resolve further the question of
the star-formation history of the Universe.

\acknowledgements{
  Funding for the creation and distribution of the SDSS Archive has
  been provided by the Alfred P.  Sloan Foundation, the Participating
  Institutions, the National Aeronautics and Space Administration, the
  National Science Foundation, the U.S. Department of Energy, the
  Japanese Monbukagakusho, and the Max Planck Society.  The SDSS Web
  site is http://www.sdss.org/.  The SDSS is managed by the
  Astrophysical Research Consortium (ARC) for the Participating
  Institutions. The Participating Institutions are The University of
  Chicago, Fermilab, the Institute for Advanced Study, the Japan
  Participation Group, The Johns Hopkins University, Los Alamos
  National Laboratory, the Max-Planck-Institute for Astronomy (MPIA),
  the Max-Planck-Institute for Astrophysics (MPA), New Mexico State
  University, Princeton University, the United States Naval
  Observatory, and the University of Washington.
  
  We acknowledge Jeff Tober (JHU) for supplying his analysis of the
  Gallego et al. (1995)
  data.

  Karl Glazebrook and Ivan Baldry acknowledge generous funding from
  the David and Lucille Packard foundation.
}


\newpage

Table A1. Cosmic Spectrum for Region A.\\
\begin{center}
\begin{tabular}{c|c|c}
\hline
Rest Wavelength  & Luminosity  & Continuum subtracted luminosity \\
\AA & $10^{30}$ W \AA$^{-1}$ Mpc$^{-3}$  & $10^{30}$ W \AA$^{-1}$ Mpc$^{-3}$ \\
\hline\hline
     6556.6 &    6.48919 &     0.30760 \\
    6558.1 &    7.10810 &     0.96345 \\
    6559.6 &    9.73218 &     3.62092 \\
    6561.1 &   15.42100 &     9.34124 \\
    6562.7 &   19.65739 &    13.61270 \\
    6564.2 &   16.94022 &    10.92763 \\
    6565.7 &   10.88128 &     4.90330 \\
    6567.2 &    7.43149 &     1.47872 \\
    6568.7 &    6.54169 &     0.62419 \\
    6570.2 &    6.34205 &     0.44839 \\
\hline
\end{tabular}
\end{center}
Note: Only a sample 10 rows are shown in a wavelength region near H$\alpha$. 
The full table covering 3764\AA--8697\AA\ is
available from the electronic edition.

\bigskip
\bigskip
\bigskip

Table A2. Cosmic Spectrum for Region B.\\
\begin{center}
\begin{tabular}{c|c|c}
\hline
Rest Wavelength  & Luminosity  & Continuum subtracted luminosity \\
\AA & $10^{30}$ W \AA$^{-1}$ Mpc$^{-3}$  & $10^{30}$ W \AA$^{-1}$ Mpc$^{-3}$ \\
\hline\hline
    3723.7 &    4.10668 &     1.17038 \\
    3724.6 &    5.54068 &     2.62494 \\
    3725.4 &    6.87602 &     3.97530 \\
    3726.3 &    7.58438 &     4.70544 \\
    3727.2 &    8.05617 &     5.19847 \\
    3728.0 &    8.50139 &     5.66404 \\
    3728.9 &    8.14010 &     5.32336 \\
    3729.7 &    6.53688 &     3.74197 \\
    3730.6 &    4.65805 &     1.85869 \\
    3731.4 &    3.45462 &     0.63813 \\
\hline
\end{tabular}
\end{center}
Note: Only a sample 10 rows are shown in a wavelength region near [OII]. 
The full table covering 3636\AA--8458\AA\ is
available from the electronic edition.

\bigskip

\end{document}